\documentclass[
aps,%
11pt,%
final,%
notitlepage,%
oneside,%
twocolumn,%
nobibnotes,%
nofootinbib,%
superscriptaddress,%
noshowpacs,%
centertags]%
{revtex4}

\begin{document}





\title{Chemical Abundances in the Globular Clusters NGC\,6229 and NGC\,6779}

\author{\firstname{D.~A.}~\surname{Khamidullina}}\email{hamidullina.dilyara@gmail.com}
\affiliation{Kazan (Volga region) Federal university, Kazan, 420008 Russia}
\author{\firstname{M.~E.}~\surname{Sharina}}
\email{sme@sao.ru} \affiliation{Special Astrophysical Observatory, Russian Academy of Sciences, Nizhnii Arkhyz, 369167 Russia}
\author{\firstname{V.~V.}~\surname{Shimansky}}\email{Slava.Shimansky@kpfu.ru}
\affiliation{Kazan (Volga region) Federal university, Kazan, 420008 Russia}
\author{\firstname{E.}~\surname{Davoust}}\email{edavoust@irap.omp.eu}
\affiliation{Research Institute in Astrophysics and Planetology, Toulouse University,\\   National Center for Scientific Research, Toulouse, 31400 France}
\affiliation{IRAP, Universit\'e de Toulouse, CNRS, Toulouse, 31400 France}


\begin{abstract}
Long-slit medium-resolution spectra of the Galactic globular
clusters (GCs) NGC\,6229 and NGC\,6779, obtained with the CARELEC
spectrograph at the 1.93-m telescope of the Haute-Provence
observatory, have been used to determine the age, helium abundance
(Y), and metallicity [Fe/H] as well as the first estimate of the
abundances of C, N, O, Mg, Ca, Ti, and Cr for these objects. We
solved this task by comparing the observed spectra and the
integrated synthetic spectra, calculated with the use of the
stellar atmosphere models with the parameters preset for the stars
from these clusters. The model mass estimates, $T_{\rm eff}$, and
$\log~g$ were derived by comparing the observed
``color--magnitude'' diagrams and the theoretical isochrones. The
summing-up of the synthetic blanketed stellar spectra was
conducted according to the Chabrier mass function. To test the
accuracy of the results, we estimated the chemical abundances,
[Fe/H], $\log~t$, and $Y$ for the NGC\,5904 and NGC\,6254
clusters, which, according to the literature, are considered to be
the closest analogues of the two GCs of our study. Using the
medium-resolution spectra from the library of Schiavon~et~al., we
obtained for these two clusters a satisfactory agreement with the
reported estimates for all the parameters within the errors. We
derived the following cluster parameters. NGC\,6229:
$[Fe/H] = -1.65$~dex, $t = 12.6$~Gyr,
$Y = 0.26$, $[\alpha/Fe] = 0.28$~dex;
NGC\,6779: $[Fe/H] = -1.9$~dex, $t =
12.6$~Gyr, $Y = 0.23$, $[\alpha/Fe] =
0.08$~dex; NGC\,5904: $[Fe/H] = -1.6$~dex, $t
= 12.6$~Gyr, $Y = 0.30$, $[\alpha/Fe] =
0.35$~dex; NGC\,6254: $[Fe/H] = -1.52$~dex,
$t = 11.2$~Gyr, $Y = 0.30$,
$[\alpha/Fe] = 0.025$~dex. The value
$[\alpha/Fe]$ denotes the average of the Ca and Mg
abundances.
\end{abstract}

\maketitle

\section{INTRODUCTION}

There are about 150 globular clusters (GCs) in our Galaxy with
different luminosities, sizes, and stellar density in the
center~\cite{harris96:Khamidullina_n}. The majority of GCs are old
($t=7$--$15$~Gyr). The clusters move with different
velocities, at different distances relative to the Galactic
center, and have different abundances. Objects with relatively
high metallicity occur closer to the Galactic center. They are
called bulge clusters. The GCs with the lowest metallicity
($-2.4<[Fe/H] < -1.5$ dex) occur in the halo of our
Galaxy. The most distant of these objects is NGC\,2419. Its
galactocentric distance is equal to $R_{\rm GC}=84$ kpc
which is 3.5~times greater than the distance $R_{\rm GC}=24$
kpc of the nearest dwarf spheroidal to the Milky Way: the Sagittarius dSph. The typical parameters of GC subsystems and
their population characteristics are described, e.g., in Borkova
and Marsakov~\cite{mars:Khamidullina_n}.

The metallicities of the majority of GCs are derived from a
comparison of stellar photometry results and theoretical
evolutionary tracks and isochrones. The brightest cluster stars
are investigated with high resolution spectrographs. By no means
all of the Galactic GCs have been studied spectroscopically. The
clusters of our study (NGC\,6229 and NGC\,6779~(M\,59)) belong to
such poorly studied objects.

\begin{table*}
\label{tab_prop:Khamidullina_n} 
\caption{The key parameters of the
studied and comparison globular clusters according
to~\cite{harris96:Khamidullina_n}: heliocentric distance ${\rm
Dist}_{\odot}$; Galactocentric distance ${\rm Dist}_{\rm GC}$,
assuming ${\rm Dist}_{\rm GC}=8$ kpc for the Sun; absolute
$V$-band magnitude in the Johnson--Cousins photometric system;
central $V$-band surface brightness; color excess; tidal radius
$r_t$; core radius~$r_{c}$; central concentration $c =
\log(r_t/r_c)$; central luminosity density logarithm}
\medskip
\begin{tabular}{l|c|c|c|c}
\hline
Parameter   & NGC\,6229 & NGC\,6779 & NGC\,5904 & NGC\,6254\\
\hline
${\rm Dist}_{\odot}$, kpc  & $30.4$ & $10.1$    & $7.5$     & $4.4$   \\
${\rm Dist}_{\rm GC}$, kpc & $29.7$ &  $6.2$    & $9.7$     & $4.6$    \\
$M_V$, mag           & $-8.05$  & $-7.38$   & $-8.81$   & $-7.48$  \\
$\mu_V$, mag arcsec$^{-2}$ & $16.99$ & $18.06$ & $16.05$ & $17.69$ \\
$E(B-V)$, mag        & $0.01$   & $0.20$    & $0.03$    & $0.28$    \\
$r_t$, pc            & $47.6$ & $25.15$   & $61.96$   & $27.49$   \\
$ r_c$, pc           & $1.15$   & $1.08$    & $0.91$    & $1.10$   \\
$c$                  & $1.61$   & $1.37$    & $1.83$    & $1.40$   \\
$\log L_{\rm cent}$, $L_{\odot}/pc^3$ & $3.41$  & $3.26$ & $3.91$ & $3.5$ \\
\hline
\end{tabular}
\end{table*}

The basic properties of the studied objects are presented in
Table~1 along with the parameters of the objects we chose
from~\cite{harris96:Khamidullina_n}; these two objects, NGC\,5904
(M\,5) and NGC\,6254 (M\,10), are their close analogues in terms of
metallicity, age, horizontal branch type, and other parameters. We
took the spectra of these GCs from Schiavon et
al.~\cite{schiavon:Khamidullina_n}. All four GCs under consideration are
massive, dynamically evolved objects with high stellar density
in the center.

NGC\,6229 is one of the most distant GCs of the outer halo. The
photometry was performed and the ``color--magnitude'' diagram
($C$--$M$) was plotted for this cluster by several authors during
1986--1991 (see references in~\cite{borissova97:Khamidullina_n}).
The first deep photometry of the cluster core and periphery,
including the full horizontal branch (HB) and the asymptotic giant
branch, was performed by Borissova et al. with the 2-m telescope
at NAO Rozhen,
Bulgaria~\cite{borissova97:Khamidullina_n,borissova99:Khamidullina_n,borissova01:Khamidullina_n}.
In these papers and also in~\cite{catellan98:Khamidullina_n}, the
HB structure and the variable stars in NGC\,6229 were
investigated; the position of the main sequence (MS) turnoff point
was also determined. Borissova et
al.~\cite{borissova99:Khamidullina_n} were the first to note the
similarity in the metallicity and age of NGC\,6229 and NGC\,5904.
A common property of these two clusters is the presence of a
considerable number of RR\,Lyrae-type variable stars. The blue
stars brighter than the MS turnoff point (blue stragglers) were
studied in~\cite{sanna12:Khamidullina_n} based on the Hubble Space
Telescope images.

NGC\,6779 is a low-metallicity globular cluster of the halo. It is
comparatively poorly studied because of its close proximity to the
Galactic plane. Hatzidimitriou et al.~\cite{hatzidimitriou:Khamidullina_n}
presented a deep ground-based photometry of this cluster, and
Sarajedini et al.~\cite{sarajedini:Khamidullina_n} and Dotter et
al.~\cite{dotter10:Khamidullina_n} performed the stellar photometry
based on the Hubble images. The authors estimated the metallicity and
age of the cluster using the Dartmouth
isochrones~\cite{dartmouth:Khamidullina_n}. Unlike NGC\,6229,
NGC\,6779 has a low metallicity (see Table~1) and an extended blue
horizontal branch.

NGC\,5904 is the closest GC to us that is located far from the
Galactic plane. That is why it is an excellent object for studying
the abundances of individual GC stars, and for plotting deep
$C$--$M$ diagrams. Its exact photometry serves for testing
the stellar evolution models. The cluster includes a large number
of variable stars. It has an enormous space velocity and a very
eccentric orbit~\cite{cud:Khamidullina_n}. Two millisecond pulsars
have been found in it, one of which is associated with a massive
neutron star~\cite{freire:Khamidullina_n}. The nature of these
objects is not clear yet. Coppola et
al.~\cite{coppola:Khamidullina_n} estimated the distance to the
cluster by the ``period--infrared magnitude'' relation for the
RR\,Lyr-type stars: $(M-m)_0=14.44 \pm 0.02$. Similar
distance estimates were obtained by approximating the evolutionary
sequences of the cluster with theoretical tracks using the Hubble
photometry
\cite{layden:Khamidullina_n,vandenberg:Khamidullina_n}.

NGC\,6254 is one of the nearby clusters of the Galactic halo for
which deep Hubble photometry is available. In particular, the
photometry results allow one to determine the percentage of
binaries as a function of distance to the center of
NGC\,6254~\cite{beccari:Khamidullina_n}. This, in turn, made it
possible to explain the low level of mass segregation in the
cluster without invoking the hypothesis of an intermediate-mass
black hole in the cluster
center~\cite{dalessandro:Khamidullina_n}. The deep photometry
performed on Hubble images and estimates of the distance and
Galactic absorption for NGC\,6254 were also presented
in~\cite{sarajedini:Khamidullina_n}.

The integrated spectra of the two nearby clusters, NGC\,5904 and
NGC\,6254, were included in the spectral library of Schiavon et
al.~\cite{schiavon:Khamidullina_n}, which was used by different
authors for numerous studies afterwards. The deep $C$--$M$
diagrams for the clusters NGC\,6779, NGC\,5904, and
NGC\,6254~\cite{layden:Khamidullina_n,sarajedini:Khamidullina_n}
were widely used for the investigation of the cluster evolutionary
status and testing the stellar population models. VandenBerg
et al.~\cite{VRmodel:Khamidullina_n} determined the age, [Fe/H],
and helium abundance index for NGC\,6779, NGC\,590, and NGC\,6254
applying their Victoria-Regina models.

Nowadays, with the development of observational and computing
techniques for the analysis of the individual stars in GCs and
also for the determination of their parameters, the method of
synthetic spectra, calculated using stellar model atmospheres, is
often used. By varying the abundances of different chemical
elements, the deviation of the theoretical spectra from the
observed ones is minimized. The used spectral range should be
large enough to include both a sufficient number of lines of the
same element with different degrees of ionization and numerous
lines of various elements. To obtain the fullest and most accurate
data, one should obviously use spectra with the highest possible
resolution. However, obtaining such spectra with a sufficiently
high $S/N$ ratio requires much observing time with large
telescopes, and is possible only for the brightest stars. The
analysis  of the integrated light of the clusters allows us to
efficiently process the data obtained with 1--2-m telescopes for
all the objects in the Galaxy, and the data from large
telescopes---for extragalactic clusters. This approach is
adopted in this paper for the four above-mentioned clusters.
It allowed us to obtain valuable data for the comparison of the
photometric and spectral stellar evolution models which is the
essential basis for understanding the properties of the stellar
populations in other galaxies.

In Section 2 the observations and their reduction techniques are
described. In Section 3 we present the method of modeling the integrated cluster spectra
in accordance with the data on the isochrones and the luminosity function,
and of determining the chemical abundances. In Section 4 the results
of the investigation are discussed, and the conclusions are formulated
in Section 5.

\section{OBSERVATIONS AND REDUCTION}

We have observed the clusters NGC\,6229 and NGC\,6779 with the
1.93-m telescope of the Haute-Provence observatory for a
comparative analysis of their stellar populations
and the characteristics derived from the study of the
$C$--$M$ diagrams. The observation log is presented in
Table~2. Long-slit medium- resolution spectra ($5.5'
\times 2''$) were obtained with the
CARELEC~\cite{coppola:Khamidullina_n} spectrograph. We used a
300~lines/mm grating, which provided a dispersion of about
$1.78$~\AA\ per pixel and a $\sim 5$~\AA (FWHM)
spectral resolution in the working spectral range of
$3700$--$6800$~\AA. Calibration exposures with HeNe lamps
were taken at the beginning and end of each night. The
observations were carried out in adverse weather conditions with
cirrus and intermittent clouds. The average seeing was
$2.5$--$3.5$". However, the lack of illumination
from the moon allowed us to obtain the CCD images of required
quality during clear sky periods. Table~2 shows that the
observations were carried out with several fixed slit positions.
In all cases the slit was oriented so as to capture not only the
central, brightest part of a GC but also the neighboring regions
highly populated with stars. Thus, the resulting 2D image
permitted us to compare the energy distributions and individual
spectral lines of the central region, unresolved into individual
objects, and the neighboring stars. As a result, we could safely
eliminate the background objects that could distort the GC
abundances from the one-dimensional sum spectrum. Note that other
authors usually use a scanning technique with a moving slit to
obtain the integrated GC spectrum (see,
e.g.,~\cite{schiavon:Khamidullina_n}). This technique enhances the
probability of background objects entering into the resulting
spectrum, especially in the case of GCs close to the Galactic
plane.

\begin{table}
\label{tab_log:Khamidullina_n} \caption{Log of spectroscopic
observations}
\medskip
\begin{tabular}{l|c|c|c}
\hline
Object     & Date       & Exposure,  & Slit \\[-5pt]
&& s &  position, deg. \\
\hline
NGC\,6229    & 09.07.10    & 600        & 0             \\
           &            & 600            & 90                    \\
NGC\,6779           & 09.07.10    & 300       & 95                    \\
     &        & 300               & 99                    \\
     & 11.07.10    & 2\,$\times$\,900                & 0                   \\
\hline
\end{tabular}
\end{table}

Primary data reduction was conducted using
MIDAS~\cite{midas:Khamidullina_n} and IRAF{\footnote {\tt
http://iraf.noao.edu}} software packages. First, we filtered out
cosmic ray hits with the {\tt filter/cosmic} code in MIDAS
environment. Hot pixels were masked. The standard procedure of
spectra reduction was then conducted: bias substraction,
flat-field correction. The dispersion relation provided an
accuracy of wavelength calibration of about $0.8$~\AA.
One-dimensional spectra were extracted with the IRAF {\tt apsum}
procedure. The final signal-to-noise ratio per pixel at the center
of the spectral range of the obtained spectra was $S/N \approx
130$. The continuum in the one-dimensional spectra was approximated
with the MIDAS {\tt filter/maximum} and {\tt filter/smooth}
(running median) functions. The analysis of the correspondence of
the observed and theoretical spectra was conducted in the Origin
6.1 graphical environment.

\section{METHOD OF DETERMINATION OF THE CHEMICAL ABUNDANCES AND EVOLUTIONARY PARAMETERS OF GLOBULAR CLUSTERS}

The method used in this paper was first presented by Sharina et
al.~\cite{sharina:Khamidullina_n}. The method applies not only to
spectral but also to photometric observational and theoretical data.
It is based on a comparison of the observed GC spectra and model
spectra derived by summing up the individual synthetic spectra of
stars with different masses, $\log~g$, and $T_{\rm eff}$. The
stellar parameters are set by the isochrone best corresponding to
the cluster $C$--$M$ diagram. The method for estimating the abundances and evolutionary parameters described in this section can be applied
to any globular cluster for which deep stellar
photometry and a long-slit spectrum containing the information on
all the stars of the cluster are available. The spectra of the dense central GC
regions are best suited for such an analysis. Our method is
applicable to spectra with a resolution of $R>2500$, a
rather high signal-to-noise ratio ($S/N>100$), and a spectral
range of no less than $1500$~\AA.

\subsection{Determination of Masses, $\log~g$, and $T_{\rm eff}$\\ for Cluster Stars}

\begin{figure*}[t!]
\setcaptionmargin{5mm}
\onelinecaptionsfalse
\includegraphics[angle=0,width=0.95\textwidth,bb=85 30 750 550]{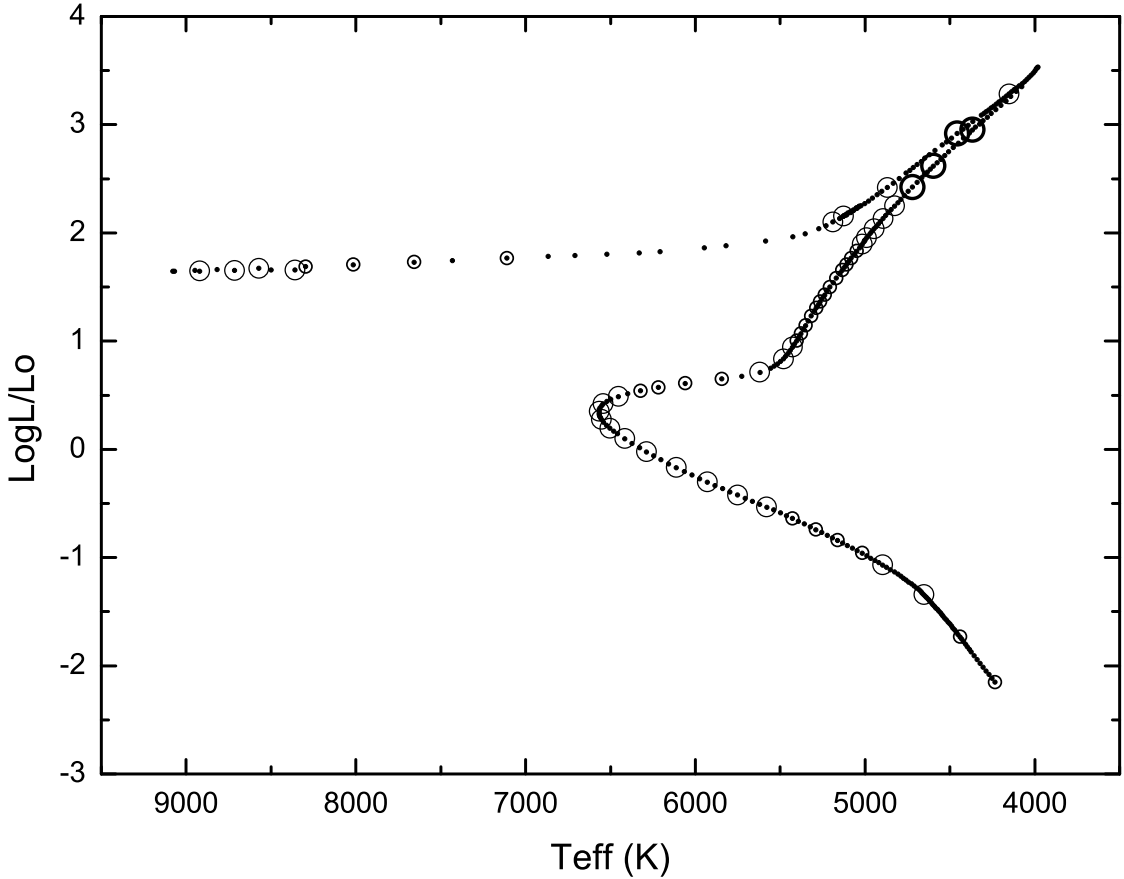}
\captionstyle{normal}
\caption{The isochrone  $\log~t=10.1$, $Y=0.26$, $Z=0.0004$
for NGC\,6229. The circles of different sizes denote the points
(in the circle centers) which contribute less
than 1\%, from 1\% to 3\%, and more than 3\% to the total flux.}
\label{fig0:Khamidullina_n}
\end{figure*}

The synthetic blanketed spectra of stellar atmospheres are calculated
according to the physical stellar parameters derived from the
comparison of the observed $C$--$M$ diagrams of the clusters and
the theoretical isochrones from Bertelli et
al.~\cite{bert1:Khamidullina_n}.\footnote {\tt
http://stev.oapd.inaf.it/YZVAR/} For our work we chose the
isochrones obtained by the Padova team. This choice was determined by the
fact that the authors included the evolutionary stages of the HB and the
asymptotic giant branch in the models and also used a wide range
of parameters: metallicity [Fe/H], age~$t$, and helium abundance
$Y$. For the clusters NGC\,6229 and NGC\,6779 we used the stellar
photometry data from Piotto et al.~\cite{piotto:Khamidullina_n}
and Sarajedini et al.~\cite{sarajedini:Khamidullina_n}.

When selecting a theoretical isochrone which best fits the cluster $C$--$M$ diagram, we vary five parameters:
age~$t$, specific helium abundance $Y$, metallicity [Fe/H], color
excess $E(B-V)$, and distance to the cluster. As initial parameters, we used data from the following references: Harris's
compiled catalog~\cite{harris96:Khamidullina_n} and Vandenberg et
al.~\cite{vandenberg:Khamidullina_n}, who determined the
parameters based on stellar photometry of Sarajedini et
al.~\cite{sarajedini:Khamidullina_n} and
Victoria-Regina~\cite{VRmodel:Khamidullina_n} models. The initial
metallicities and ages for a number of clusters were adopted
from Dotter et
al.~\cite{dotter10:Khamidullina_n}, whose results were derived with the use of
isochrones from Dartmouth~\cite{dartmouth:Khamidullina_n}.

Isochrone selection is carried out in the following way.
\begin{list}{}{
\setlength\leftmargin{4mm} \setlength\topsep{2mm}
\setlength\parsep{0mm} \setlength\itemsep{2mm} }

 \item (1) The position of the basic evolutionary sequences: main branch,
subgiant branch, red giant branch, and horizontal branch, is
computed from the photometric data as a running median with a step
of $0.1$' along two coordinates (color and magnitude). Stars with
magnitude measurement errors larger than $0.2$' should
previously have been excluded from the photometry table. Those are the faint
objects. Usually, the percentage of their detection on Hubble
images does not exceed 50$\%$. Consequently, {\it the fiducial sequence} of the $C$--$M$ diagram is plotted, which describes the
position of the basic evolutionary sequences of a cluster.

 \item (2) Next, the theoretical isochrone minimally deviating
from the fiducial sequence is selected. This is done by $\chi$-square
minimization:
$$ \chi^2  = \displaystyle\sum_{i=1}^{N} \left[
\displaystyle\frac{I_o - I_m}{\sigma_{I_o}} \right],
$$
where$I_o$,~$ \sigma_{I_o}$ are the magnitudes of the fiducial and their errors, and $I_m$ are the corresponding isochrone
values. Note that we do not interpret theoretical isochrones but
merely use the available set~\cite{bert1:Khamidullina_n}.
\end{list}

Variations of the Galactic absorption in the direction of a GC and
the distance to a cluster do not influence the inclination, form,
and relative position of different evolutionary sequences, but
shift the $C$--$M$ diagram as a whole. Variations of
[Fe/H], $t$, or $Y$, on the other hand, influence the position and
form of the isochrones (see, e.g., the explanation
in~\cite{bert1:Khamidullina_n}). The evaluation of the basic
evolutionary parameters from the $C$--$M$ diagram is conducted taking into account these variations. In the following six paragraphs, we
explain the principles of determining [Fe/H], $t$, and $Y$.

{\it Selection of [Fe/H]} ($t$, $Y$ are constant) is performed by
the inclination of the red giant
branch~\cite{salaris02:Khamidullina_n} as the basic criterion. The inclination of the branch increases with the increase of [Fe/H], and the luminosity
of the stars at the top of the red giant branch falls more and more rapidly at $[Fe/H] \ge -0.8$~dex. Moreover, with increasing metallicity, the whole isochrone shifts toward the red
region, the luminosity of the stars in it decreases, and fewer hot blue
stars appear on the HB.

{\it Selection of the age value} ([Fe/H] and $Y$ are constant) is done
using the temperature and luminosity of the MS turnoff stars. With increasing age, the MS turnoff point position shifts toward cooler
and fainter stars and is additionally dependent on the GC chemical
abundance. The luminosity of HB stars is mainly determined by
metallicity and helium abundance. There are two basic and common
empirical approaches to relative age estimation for GCs: a
vertical
one~\cite{iben71:Khamidullina_n,iben91:Khamidullina_n,VDBerg90v:Khamidullina_n},
which takes into account the difference in the luminosities of the MS turnoff
point and the zero age HB ($\delta V=V_{\rm TO} - V_{\rm
ZAHB}$), and a horizontal one~\cite{VDBerg90g:Khamidullina_n},
based on the difference in temperature (color) between the MS
turnoff point and the base of the red giant branch
$\delta(B-V)=(B-V)_{\rm RGB}-(B-V)_{\rm TO}$.

In~\cite{iben91:Khamidullina_n}, an approximate ratio between
$\delta V$, the abundances of helium $Y$ and heavy elements $Z$ on
the one hand, and the logarithm of age on the other hand is given.
As a consequence of the strong dependence of the HB star
luminosities on $Y$, and the MS turnoff point on $Z$, the vertical
method provides only the {\it relative} ages for GCs of similar
metallicity and $Y$. For the estimation of {\it absolute} ages,
the exact values of the helium mass fraction are needed.

The essence of the horizontal method is in measuring the
color difference between the MS turnoff point, sensitive to age,
and the red giant branch, sensitive to metallicity. This method is
also relative. The compared clusters should have the same
metallicities and $[\alpha/Fe]$.

It is necessary to note that both the vertical and horizontal
methods do not require knowledge of absolute distances to the
observed objects and of the light absorption in their direction.
There are numerous modifications of these two approaches (see,
e.g.,~\cite{carney:Khamidullina_n,roediger:Khamidullina_n}.
The results of age estimation for the same cluster using different
methods vary greatly. Modern researchers use a maximum likelihood
method for isochrone selection for the $C$--$M$ diagram of a
cluster as a whole. We use a similar method. Estimation of the MS
turnoff point magnitude is rather difficult and often leads to an
age inaccuracy of about 10\%. That is why the approximation by an
isochrone of the $C$--$M$ diagram as a whole with all its bends is
much more precise. As we use not only the photometric data
characterizing the stellar luminosities and temperatures but also
spectroscopic data on the detailed abundances, further development
of our method could allow us to estimate the absolute ages of the
clusters.

{\it  Variation of $Y$} influences mainly the luminosity and
temperature of HB stars. Old GCs have many blue stars on the HB
and often a sequence of low-luminosity stars with temperatures
higher than those of the HB stars~\cite{cruz:Khamidullina_n}.

After the selection of [Fe/H], $t$, and $Y$ using the
$C$--$M$ diagram, the spectrum with model parameters
corresponding to the selected isochrone is calculated. Further
refinement of the values [Fe/H], $t$, and $Y$ is obtained with
the use of spectroscopic data (see detailed explanation in
paragraph~3.2). The long slit is adjusted to the cluster's center.
Consequently, the spectrum of compact GCs includes the information
on all their stars. Provided that there is no
contribution from bright background stars, the integrated
spectrum allows us to match the data derived from the $C$--$M$
diagram with the results of synthetic spectra obtained with stellar atmosphere models. For calculating the spectra, the
number of isochrone points was optimized by eliminating the
points contributing less than
$0.5\%$ to the total luminosity of a cluster. To estimate the contribution of individual points to the
total light of a cluster, the flux in the continuum at the wavelength
$\lambda = 5000$~\AA, calculated for the model
atmosphere with the parameters of a given point ($T_{\rm eff}$,
$\log~g$ and [Fe/H]), is multiplied by the squared radius of the
star, the weight of the point in the full mass interval, and the current value
of the mass function. The flux derived this way is further
divided by the total flux from all the points of an isochrone.
Figure~\ref{fig0:Khamidullina_n} shows isochrone points which contribute
to the total light of NGC\,6229 by less than 1\%, from
1\% to 3\% and more than 3\% in circles of different sizes.

\begin{figure*}[tbp!!!]
 \vspace{2mm}
\begin{minipage}{0.49\linewidth}
\setcaptionmargin{5mm} \onelinecaptionsfalse
\includegraphics[width=\columnwidth,bb=0 0 794 590,clip]{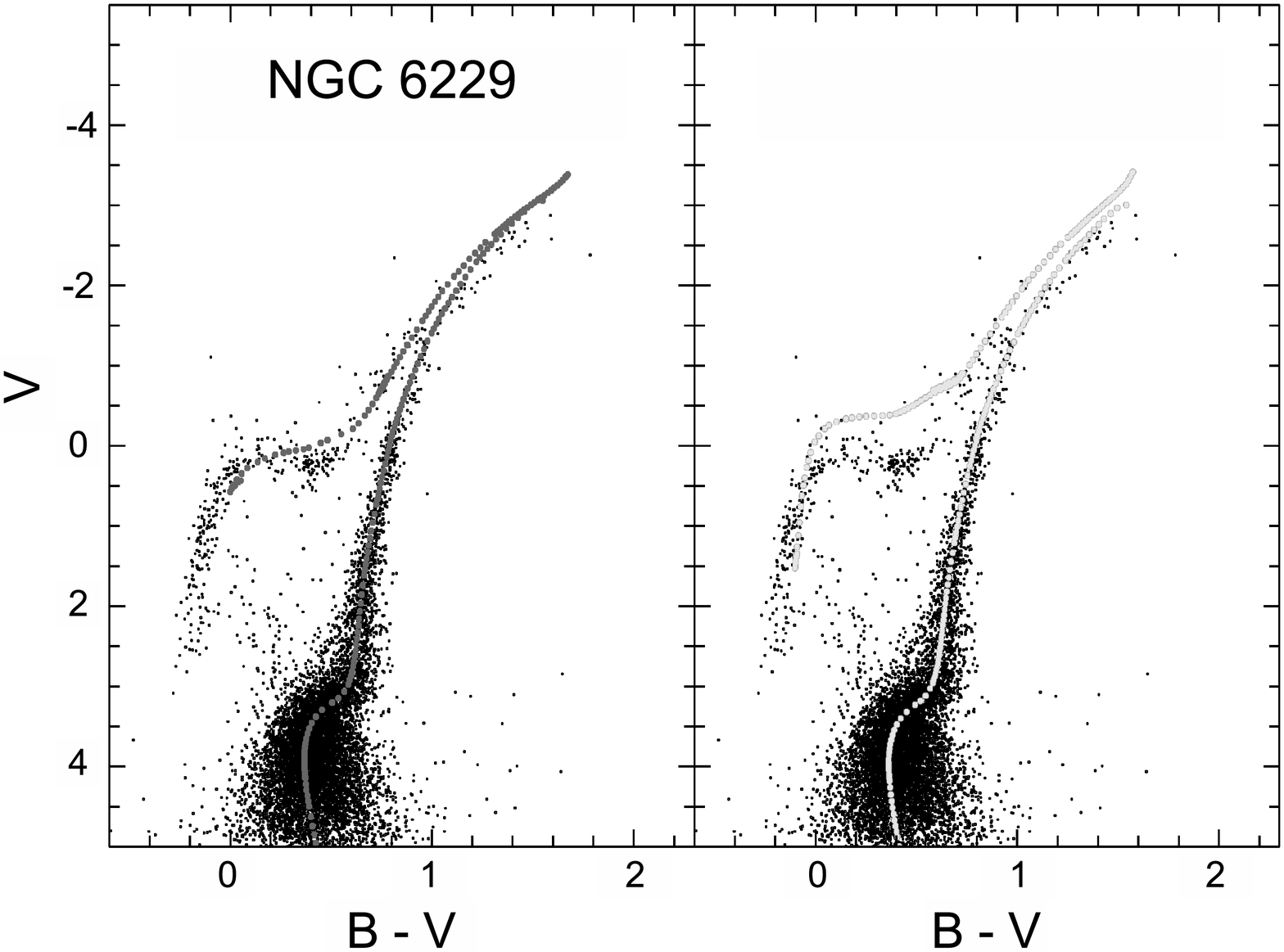}
\captionstyle{normal}
\vspace{-5mm}
\caption{Comparison of the $C$--$M$ diagram $V$--$(B-V)$ for
NGC\,6229~\cite{piotto:Khamidullina_n} with the isochrones from
Bertelli et al.~\cite{bert1:Khamidullina_n}. The left panel shows
the isochrone $\log~t=10.1$, $Y=0.26$,
$Z=0.0004$ used in the GC spectrum modelling. On the right,
the isochrone $\log~t=10.1$, $Y=0.3$,
$Z=0.0004$ (see Section 3.1) is shown for comparison.}
\label{fig1:Khamidullina_n}
\end{minipage}
\begin{minipage}{0.49\linewidth}
\setcaptionmargin{5mm} \onelinecaptionsfalse
\includegraphics[width=\columnwidth,bb=0 0 794 590,clip]{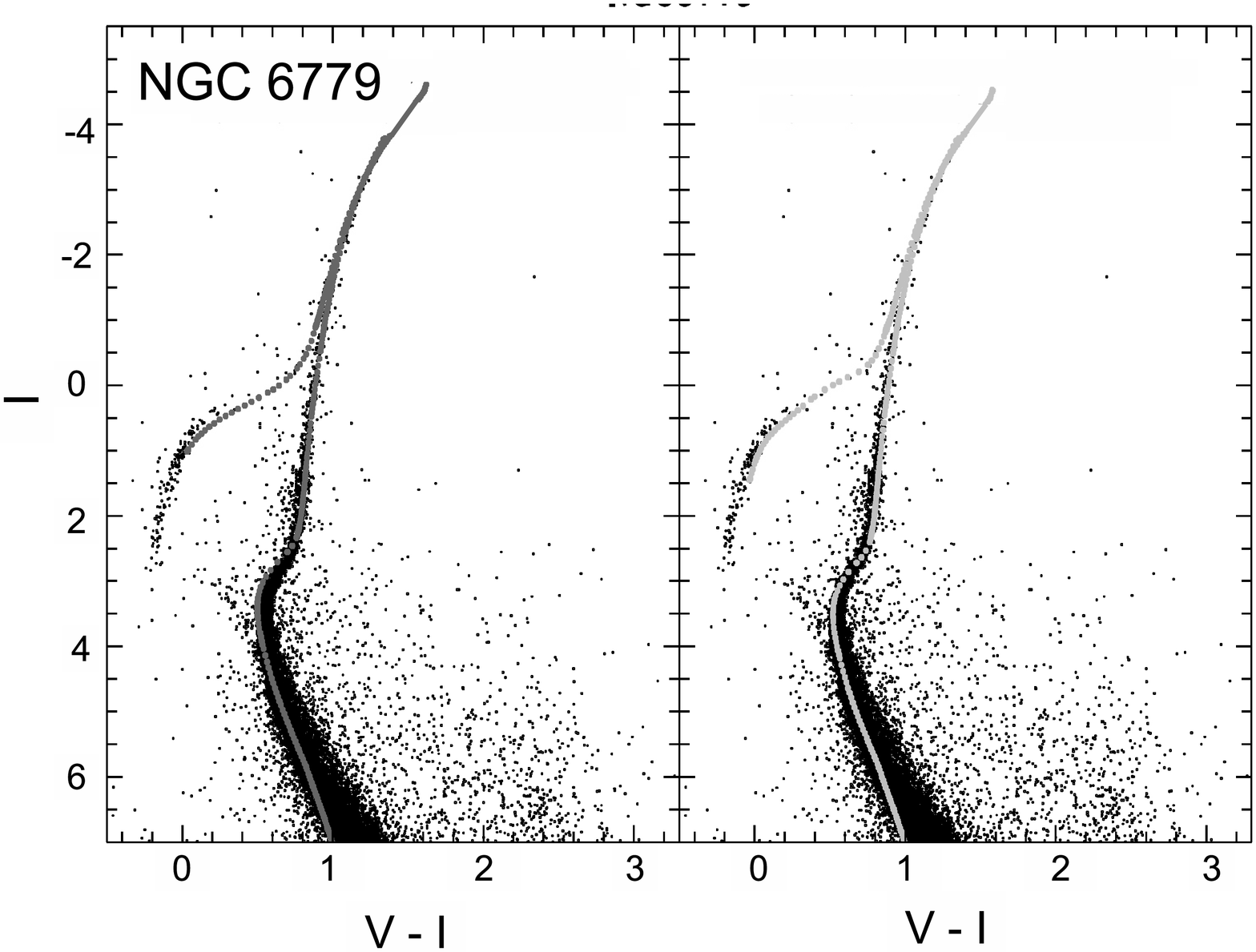}
\caption{Comparison of the $C$--$M$ diagram $I$--$(V-I)$ for
NGC\,6779~\cite{sarajedini:Khamidullina_n} with the isochrones
from~\cite{bert1:Khamidullina_n}. The left panel shows the
isochrone $\log~t=10.1$, $Y=0.23$, $Z=0.0001$ used in the GC
spectrum modelling. On the right, the isochrone $\log~t=10.15$,
$Y=0.23$, $Z=0.0001$ is shown for comparison.}
\label{fig2:Khamidullina_n}
\end{minipage}
\end{figure*}

\begin{figure*}[tbp!!!]
\begin{minipage}{0.49\linewidth}
\setcaptionmargin{5mm} \onelinecaptionsfalse
\includegraphics[width=\columnwidth,bb=0 0 794 590,clip]{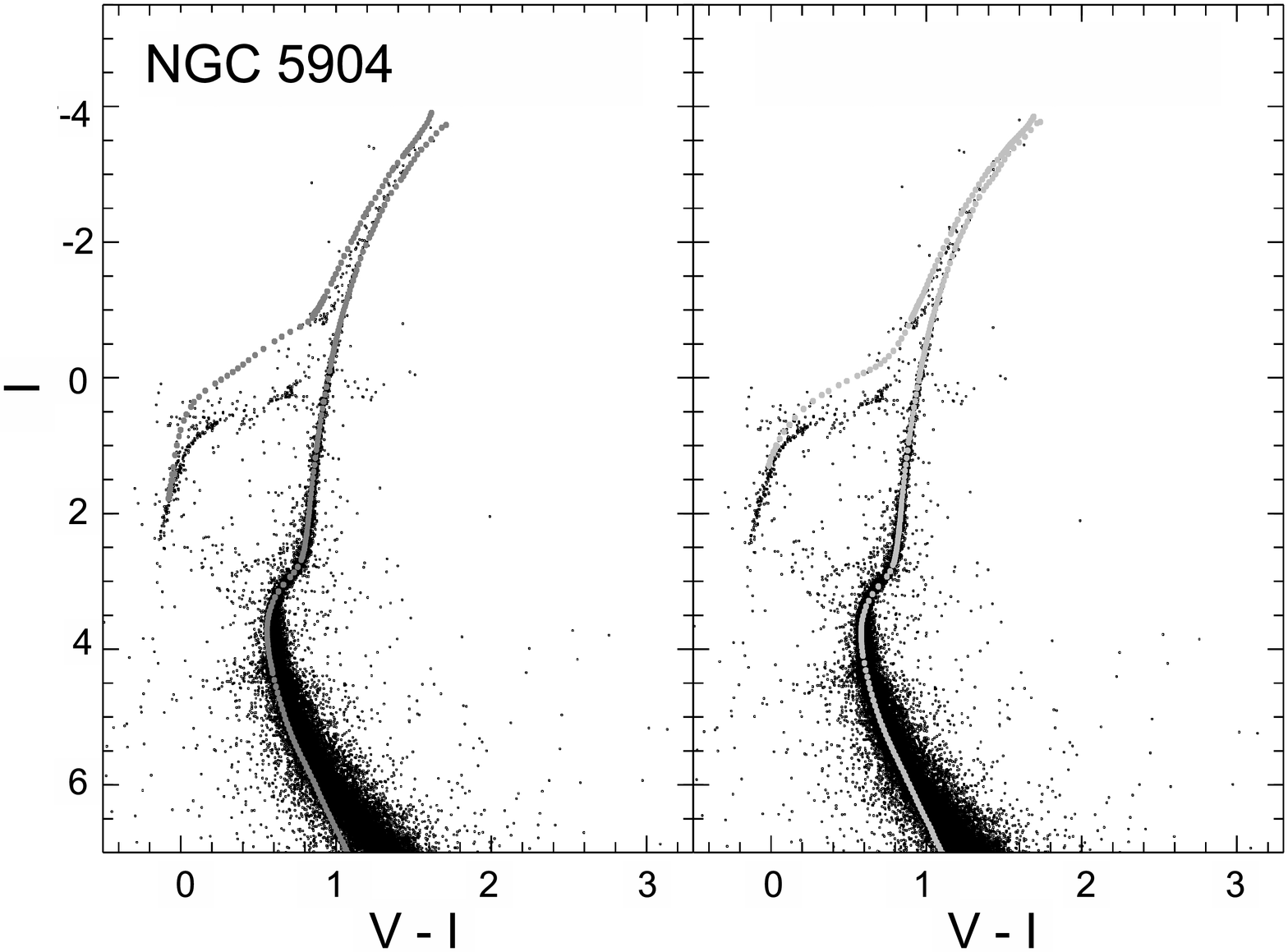}
\captionstyle{normal}
\caption{Comparison of the $C$--$M$ diagram $I$--$(V-I)$ for
NGC\,5904~\cite{piotto:Khamidullina_n} with the isochrones
from~\cite{bert1:Khamidullina_n}. The left panel shows the
isochrone $\log~t=10.1$, $Y=0.30$, $Z=0.001$
used in the GC spectrum modelling. On the right, the isochrone
$\log~t=10.15$, $Y=0.26$, $Z=0.001$ (see
Section 3.1) is shown for comparison.} \label{fig3:Khamidullina_n}
\end{minipage}
\begin{minipage}{0.49\linewidth}
\setcaptionmargin{5mm} \onelinecaptionsfalse
\includegraphics[width=\columnwidth,bb=0 0 794 590,clip]{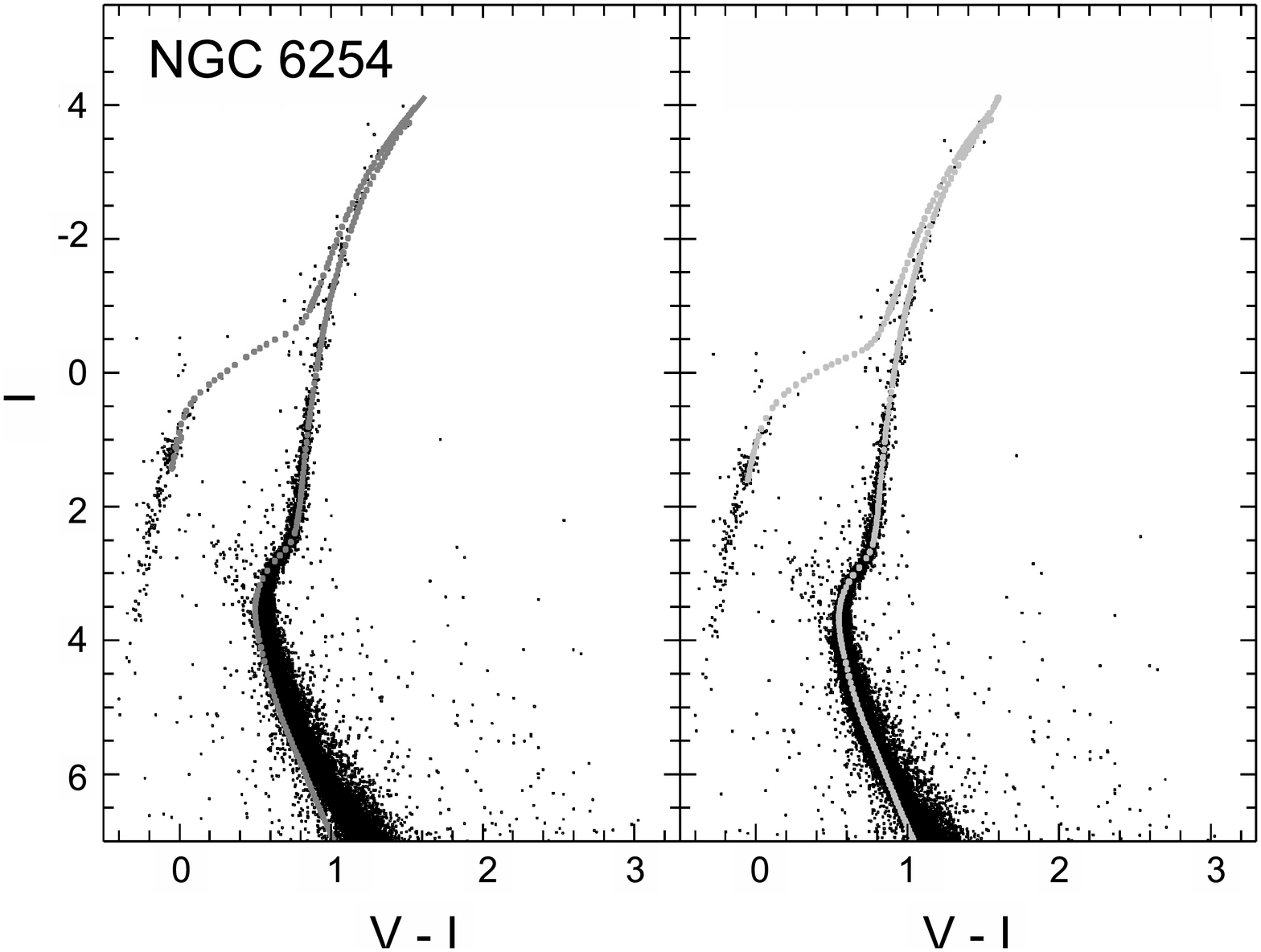}
\captionstyle{normal}
\caption{Comparison of the $C$--$M$ diagram $I$--$(V-I)$ for
NGC\,6254~\cite{sarajedini:Khamidullina_n} with the isochrones
from~\cite{bert1:Khamidullina_n}. The left panel shows the
isochrone $\log~t=10.05$, $Y=0.30$,
$Z=0.0004$, used in the GC spectrum modelling. On the
right, the isochrone $\log~t=10.15$, $Y=0.26$,
$Z=0.0004$ is shown for comparison.}
\label{fig4:Khamidullina_n}
\end{minipage}
\end{figure*}

\begin{figure*}[tbp!!!]
\setcaptionmargin{5mm}
\onelinecaptionsfalse
\includegraphics[angle=-90,width=0.95\textwidth]{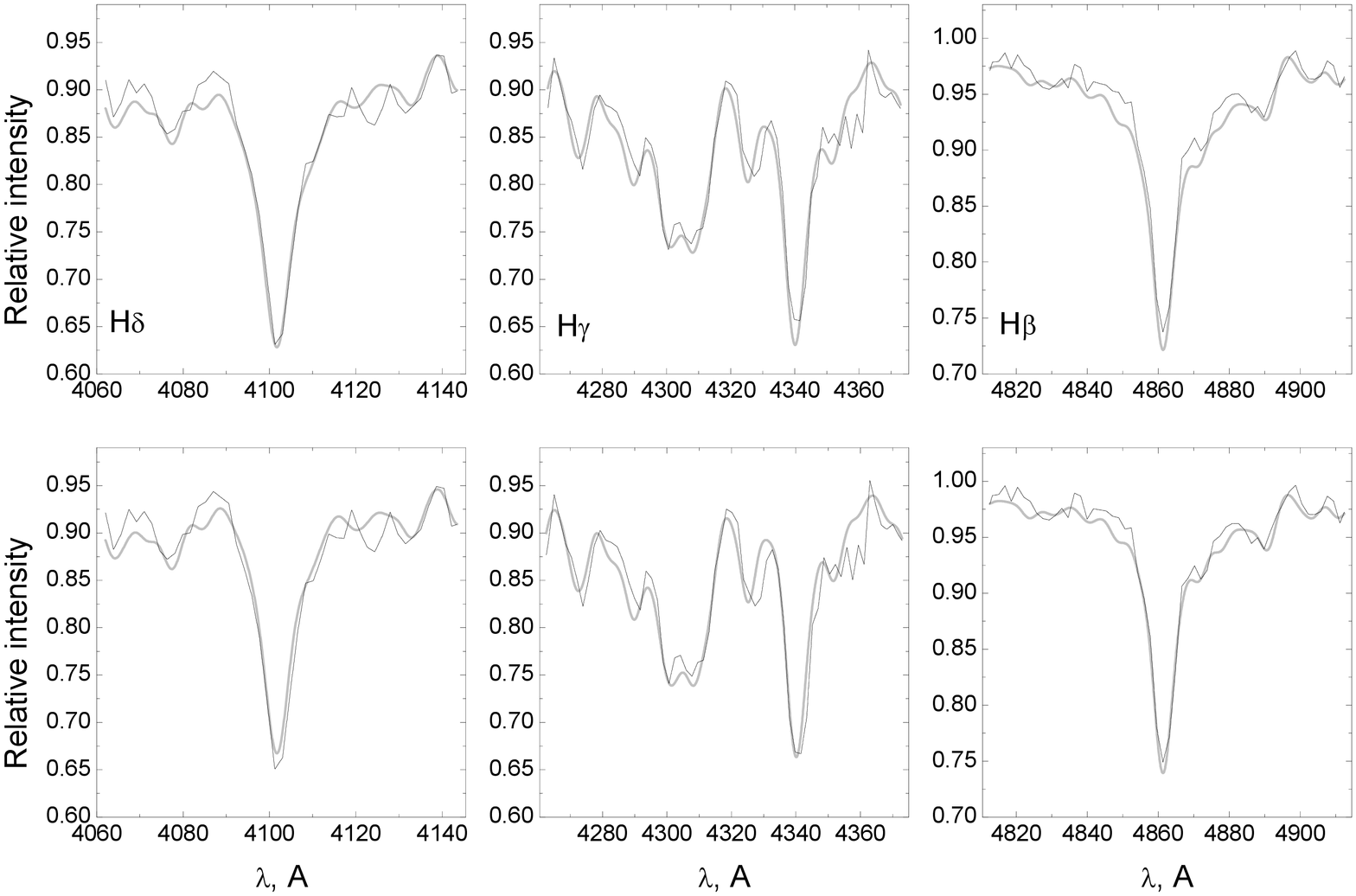} 
\captionstyle{normal}
\caption{The fit of the isochrone parameters to the observed
spectrum of NGC\,6229 (the profiles of the
H$\beta$,~H$\gamma$,~H$\delta$ lines). The three upper panels show
the applied model $\log~t=10.1$, $Y=0.26$,
$Z=0.0004$ (see the corresponding $C$--$M$ diagram on the
left of Fig.~\ref{fig1:Khamidullina_n}). The three lower panels
show the model $\log~t=10.1$, $Y=0.30$,
$Z=0.0004$ (the right-hand side of
Fig.~\ref{fig1:Khamidullina_n}) for comparison.}
\label{fig5:Khamidullina_n}
\end{figure*}

\begin{figure*}[tbp!!!]
\setcaptionmargin{5mm}
\onelinecaptionsfalse
\includegraphics[angle=-90,width=0.95\textwidth]{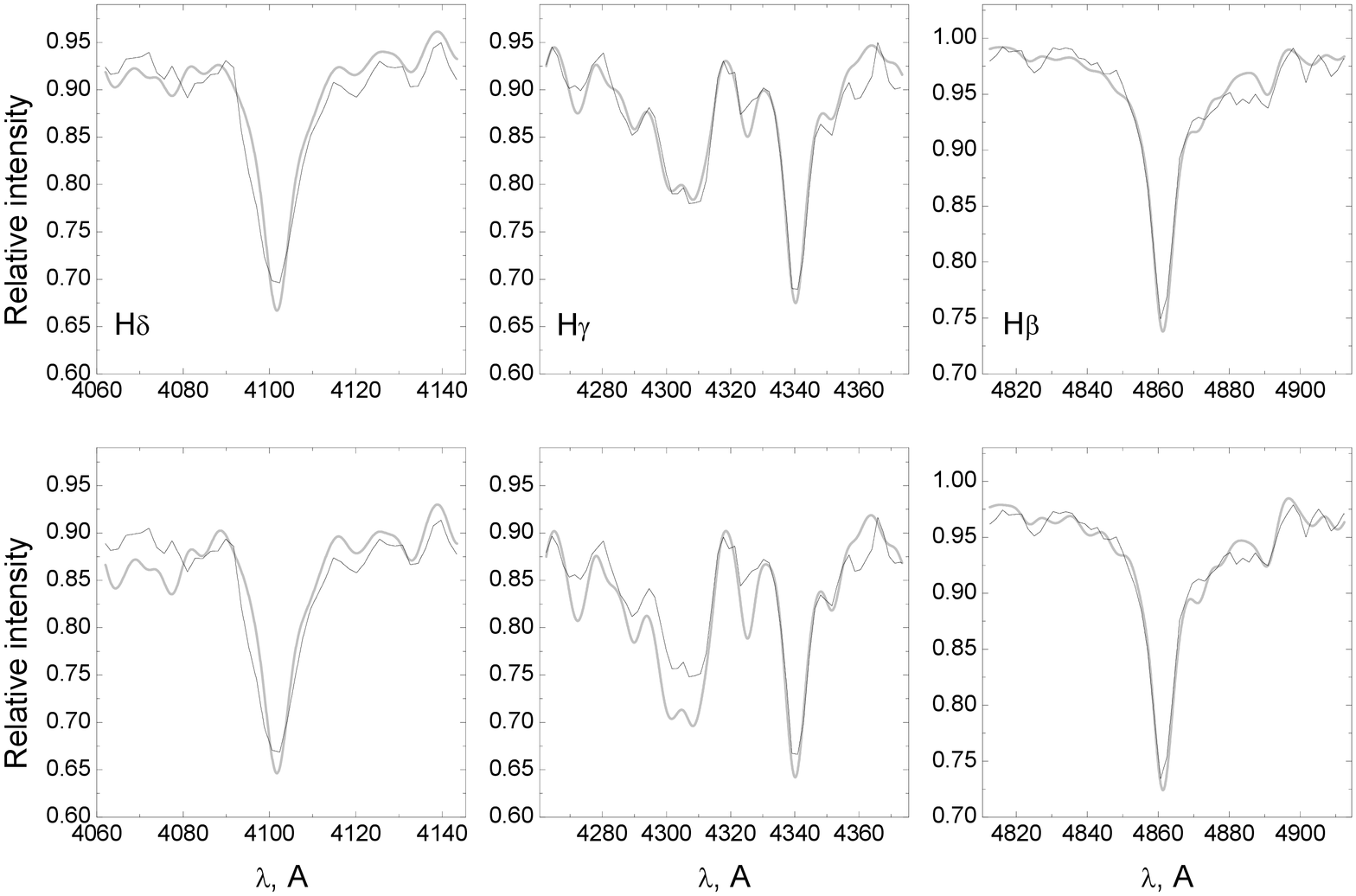}
\captionstyle{normal}
\caption{The fit of the isochrone parameters to the observed
spectrum of NGC\,6779 (the profiles of the
H$\beta$,~H$\gamma$,~H$\delta$ lines). The three upper panels show the
applied model $\log~t=10.1$, $Y=0.23$,
$Z=0.0001$ (see the corresponding $C$--$M$ diagram on the left of
Fig.~\ref{fig2:Khamidullina_n}). The three lower panels
show the model $\log~t=10.15$, $Y=0.23$,
$Z=0.0001$ (the right-hand side of Fig.~\ref{fig2:Khamidullina_n})
for comparison.} \label{fig5_1:Khamidullina_n}
\end{figure*}

\begin{figure*}[tbp!!!]
\setcaptionmargin{5mm}
\onelinecaptionsfalse
\includegraphics[angle=-90,width=0.95\textwidth]{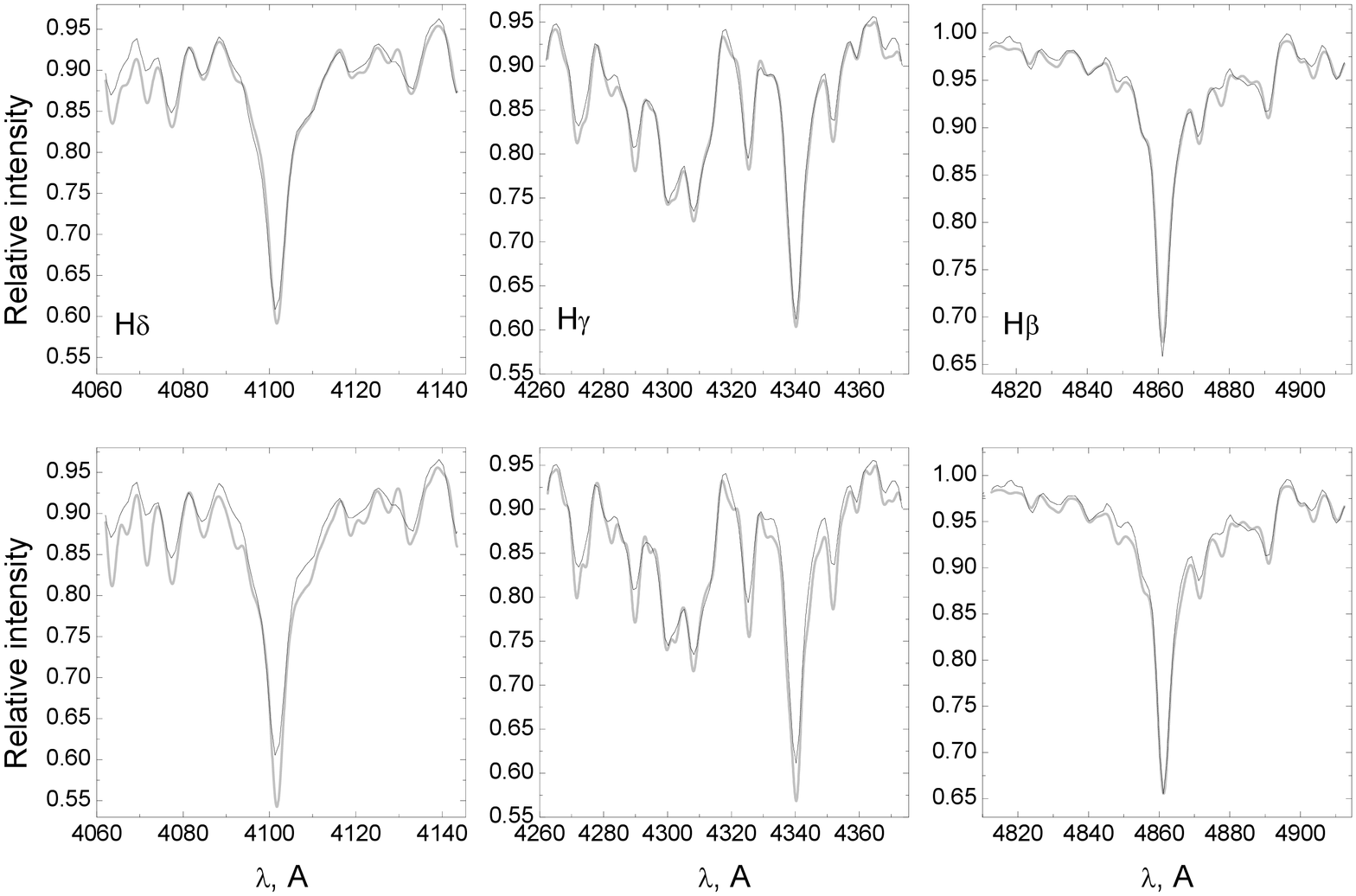}
\captionstyle{normal}
\caption{The fit of the isochrone parameters to the observed
spectrum of NGC\,5904 (the profiles of the
H$\beta$,~H$\gamma$,~H$\delta$ lines). The three upper panels show
the applied model $\log~t=10.1$, $Y=0.30$,
$Z=0.001$ (see the corresponding $C$--$M$ diagram on the
left of Fig.~\ref{fig3:Khamidullina_n}). The three lower panels
show the model $\log~t=10.15$, $Y=0.26$,
$Z=0.001$ (the right-hand side of
Fig.~\ref{fig3:Khamidullina_n}) for comparison.}
\label{fig5_2:Khamidullina_n}
\end{figure*}

\begin{figure*}[tbp!!!]
\setcaptionmargin{5mm}
\onelinecaptionsfalse
\includegraphics[angle=-90,width=0.95\textwidth]{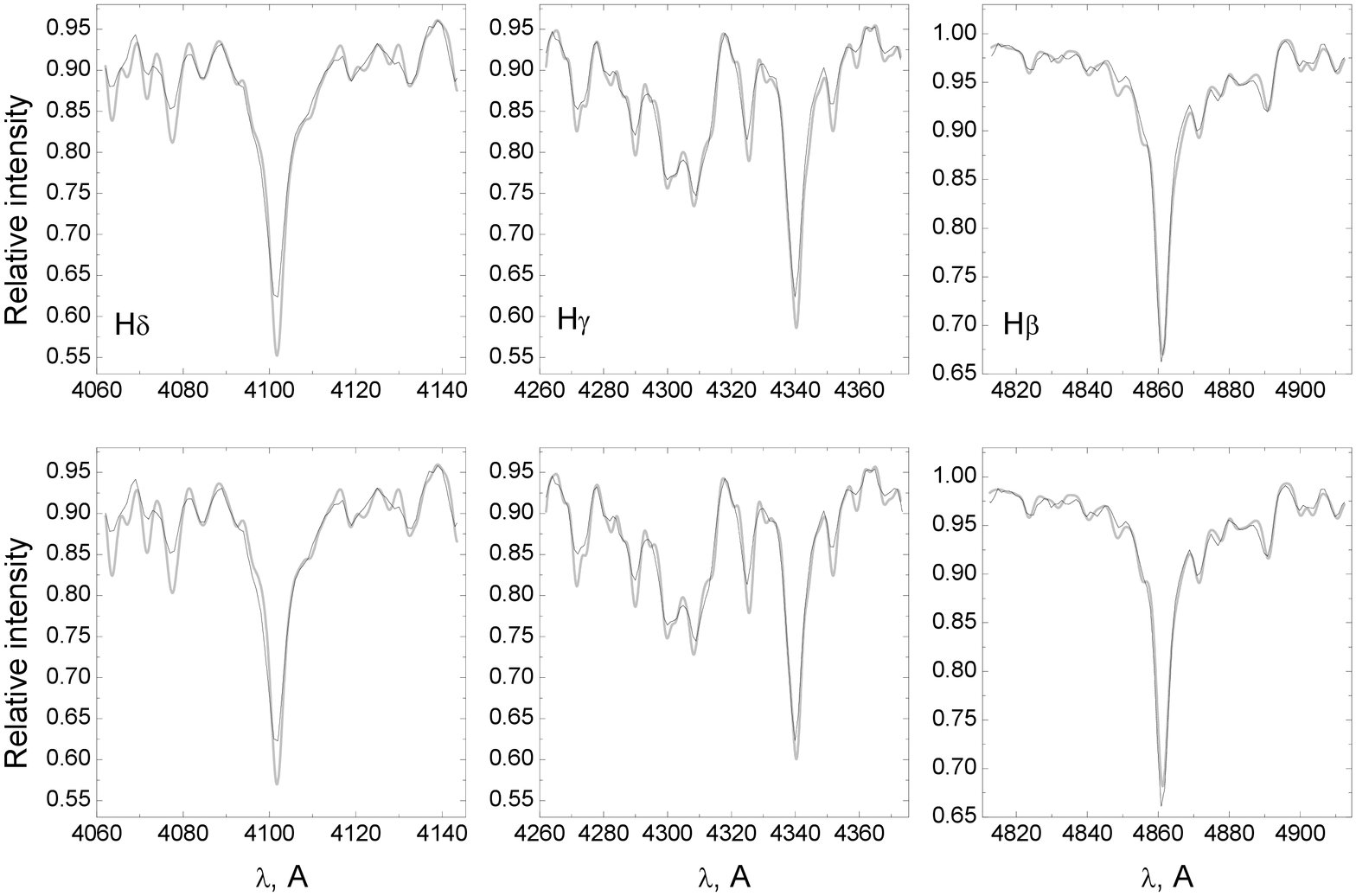}
\captionstyle{normal}
\caption{The fit of the isochrone parameters to the observed
spectrum of NGC\,6254 (the profiles of the
H$\beta$,~H$\gamma$,~H$\delta$ lines). The three upper panels show the
applied model $\log~t=10.05$, $Y=0.30$,
$Z=0.0004$ (see the corresponding $C$--$M$ diagram on the left of
Fig.~\ref{fig4:Khamidullina_n}). The three lower panels
show the model $\log~t=10.15$, $Y=0.26$,
$Z=0.0004$ (the right-hand side of Fig.~\ref{fig4:Khamidullina_n})
for comparison.} \label{fig6:Khamidullina_n}
\end{figure*}

The derived results of isochrone selection using the $C$--$M$
diagram and the spectra for the globular clusters under study are
shown in the left panels of
Figs.~\ref{fig1:Khamidullina_n}--\ref{fig4:Khamidullina_n}.
On the right-hand side, isochrone positions with varied parameters
are presented. For NGC\,6229 the variation of the isochrone
position with increasing $Y$ is shown, for NGC\,6779---with
increasing age, and for NGC\,5904 and NGC\,6254---with the
variation of $Y$ and age. For the last two clusters, isochrones
with parameters close to those estimated
in~\cite{VDBerg90g:Khamidullina_n} are presented. Distance and
$E(B-V)$ were changed according to variations of the above
mentioned parameters. All GCs have certain stars and sequences
which cannot be described by iso\-chrones. They are hot blue HB
stars, blue stragglers, certain bright stars deviating from the
red giant branch (background objects or asymptotic giant branch
stars). In Section~\ref{model_atm:Khamidullina_n} we discuss the
significance of the listed deviations during the analysis of the
correspondence of theoretical spectra calculated with the
specified isochrones to the observed ones.
Figs.~\ref{fig5:Khamidullina_n}--\ref{fig6:Khamidullina_n}
show the spectral regions with hydrogen lines calculated with the
parameters of isochrones which are used in
Figs.~\ref{fig1:Khamidullina_n}--\ref{fig4:Khamidullina_n}. As the
final values of [Fe/H], $t$ and $Y$, we adopted those that best
describe the observed spectra.

The results of isochrone selection for the thoroughly studied
clusters NGC\,5904 and NGC\,6254
\linebreak(Figs.~\ref{fig3:Khamidullina_n}~and~\ref{fig4:Khamidullina_n})
agree with the quantities from the literature
(Table~3)~\cite{harris96:Khamidullina_n,vandenberg:Khamidullina_n}.
The differences in age estimates do not exceed 1 Gyr. However, the
[Fe/H] value is regularly underestimated by
$0.1$--$0.3$~dex for all the clusters except NGC\,6779. For
the two tested clusters, the helium abundance was found to be higher
than in the photometric results of VandenBerg et
al.~\cite{vandenberg:Khamidullina_n}. The $\alpha$-element
abundance for NGC\,6254 that we measured turned out to be
$0.2$--$0.3$ dex lower than the value reported in the
literature. On the whole, our errors of the determination of the
main cluster parameters do not exceed 20\%.

\begin{table*}{}
\label{tab_prop1:Khamidullina_n}
 \caption{Comparison of the values taken from the
literature~\cite{harris96:Khamidullina_n,vandenberg:Khamidullina_n}
and those determined by us using photometric and spectroscopic methods:
[Fe/H] (in dex), helium abundance $Y$, age $t$ (in Gyr), $\alpha$-element abundance
$[\alpha/Fe]=\left([Mg/Fe]+[Ca/Fe]\right)/2$
(in dex), and average microturbulence velocity for each cluster
$\xi_{\rm turb}$ in km\,s$^{-1}$. The parameter ${\rm
HBR}=(B-R)/(B+V+R)$~\cite{harris96:Khamidullina_n} characterizes
the number of stars in the blue and red regions of the horizontal
branch and in the Hertzsprung gap}
\medskip
\begin{tabular}{c|c|c|c|c|c}
\hline
Parameter & NGC\,6229 & NGC\,6779 & NGC\,5904 & NGC\,6254 & Ref.\\
\hline
[Fe/H], dex                             & $-1.43$ & $-1.94$ & $-1.27$ & $-1.52$ & [1]\\
${\rm HBR} $                             & $0.24 $ & $ 0.98$ & $0.31 $ & $0.98 $ & [1]\\
$[\alpha/Fe] $              & --    & --    & $0.21 $ & $0.26 $ & [37] \\
$[\alpha/Fe] $              & --    & --    & $0.31 $ & $0.39 $ & [31] \\
$[Fe/H]$                    & --    & $-2.00$ & $-1.33$ & $-1.57$ & [17]\\
$ t $                              & --    & $12.75\pm0.50$ & $11.50\pm0.25$ & $11.75\pm0.38$ & [17]\\
$Y $                               & --    & $0.25 $ & $0.25 $ & $0.25 $ & [17]\\
$[Fe/H]_{\rm isochr.}$      & $-1.74$ & $-2.35$ & $-1.33$ & $-1.74$ & ours\\
$ t _{\rm isochr.+spectra}$        & $12.6 $ & $12.6 $ & $12.6 $ & $11.2 $ & ours\\
$Y_{\rm isochr.+spectra}$          & $0.26 $ & $0.23 $ & $0.30 $ & $0.30 $ & ours\\
$[Fe/H]_{\rm spectra}$      & $-1.65$ & $-1.9 $ & $-1.6 $ & $-1.75$ & ours \\
$[\alpha/Fe]_{\rm spectra}$ & $0.28 $ & $0.08 $ & $0.35 $ & $0.025$ & ours \\
$\xi_{\rm turb}$     & $2.1  $ & $2.1  $ & $2.1  $ & $1.8  $ & ours \\
\hline
\end{tabular}
\end{table*}

\begin{table*}{}
\setcaptionmargin{0mm} \onelinecaptionsfalse \captionstyle{normal}
\label{tab_abund1:Khamidullina_n}
 \caption{Derived abundances [X/Fe] (in dex) and their dispersion $\sigma$ (in dex)
for NGC\,6229 and NGC\,5904. The corresponding values from \cite{roediger:Khamidullina_n}
are listed in the last column.}
 \medskip
\begin{tabular}{|l|rc|rc|rc|}
\hline
  & NGC& 6229 & NGC& 5904 & NGC& 5904\cite{roediger:Khamidullina_n} \\
\hline
Elem. & [X/Fe] & $\sigma$ & [X/Fe] & $\sigma$ & [X/Fe] & $\sigma$ \\
\hline
C    & $-0.05$ & $0.15$ &$-0.20$&$0.10$& $-0.48$ &$ 0.26$ \\
N    & $ 0.35$ & $0.25$ &$ 0.20$&$0.20$& $0.68 $& $0.59 $\\
O    & $ 0.35$ & $0.30$ &$ 0.20$&$0.30$& $0.15 $& $0.27 $\\
Na   & $ 0.25$ & $0.20$ &$ 0.20$&$0.15$& $0.19 $& $0.26 $\\
Mg   & $ 0.15$ & $0.15$ &$ 0.40$&$0.15$& $0.33 $& $0.10 $\\
Ca   & $ 0.40$ & $0.15$ &$ 0.20$&$0.10$& $0.28 $& $0.11 $\\
Ti   & $ 0.40$ & $0.25$ &$ 0.20$&$0.20$& $0.22 $& $0.10 $\\
Cr   & $ 0.15$ & $0.20$ &$ 0.10$&$0.15$& $-0.08$ &$ 0.19$ \\
\hline
\end{tabular}
\end{table*}

\begin{table*}{}
\setcaptionmargin{0mm} \onelinecaptionsfalse \captionstyle{normal}
\label{tab_abund2:Khamidullina_n}
 \caption{Derived abundances [X/Fe] (in dex) and their dispersion $\sigma$ (in dex)
for NGC\,6779 and NGC\,6254. The corresponding values from \cite{pritzl05:Khamidullina_n}
are listed in the last column.}
\medskip
 \begin{tabular}{|l|rc|rc|rc|}
\hline
      & NGC& 6779         & NGC& 6254         & NGC& 6254\cite{pritzl05:Khamidullina_n}\\
\hline
Elem. & [X/Fe] & $\sigma$ & [X/Fe] & $\sigma$ & [X/Fe] & $\sigma$ \\
\hline
C    & $-0.15$ & $0.15$ & $-0.15$&$0.10$&$-0.77$&$0.37$     \\
N    & $-0.10$ & $0.25$ & $0.25 $&$0.20$&$1.01 $&$0.45$         \\
O    & $-0.20$ & $0.30$ & $0.40 $&$0.30$&$0.23 $&$0.24$      \\
Na   & $-0.20$ & $0.20$ & $0.25 $&$0.15$&$0.17 $&$0.27$         \\
Mg   & $-0.10$ & $0.15$ & $0.05 $&$0.15$&$0.44 $&$0.13$    \\
Ca   & $ 0.25$ & $0.15$ & $0.00 $&$0.10$&$0.33 $&$0.11$    \\
Ti   & $ 0.10$ & $0.25$ & $0.45 $&$0.20$&$0.26 $&$0.12$   \\
Cr   & $-0.10$ & $0.20$ & $-0.09$&$0.15$&$0.001$&$0.15$    \\
\hline
\end{tabular}
\end{table*}

However, we note the common problem (in the literature and ours)
of describing the cluster $C$--$M$ diagram using the
evolutionary tracks and isochrones. Actually, none of the
presented isochrones describes the $C$--$M$ diagram of the cluster
in all details. We selected the isochrones that provide the best
agreement between the observed and theoretical spectra calculated
using the distribution of stars by mass, $\log~g$, and $T_{\rm
eff}$ corresponding to these isochrones. Note that besides the
above-mentioned particular stars with noticeable deviations from
the isochrone, in some other cases (see, e.g.,
Fig.~\ref{fig3:Khamidullina_n}), the spectrum analysis implies a
small shift of the MS turnoff point toward higher temperatures or
a change of the HB position. For NGC\,5904 such changes may be
caused by the increased percentage of hot blue HB stars, the EHB
(Extended Horizontal Branch), and/or blue stragglers caught in the
spectrograph slit.

\subsection{Modelling of Stellar Spectra}
\label{model_atm:Khamidullina_n}

\begin{figure*}[tbp!!!]
\setcaptionmargin{5mm}
\onelinecaptionsfalse
\includegraphics[angle=0,width=0.9\textwidth,bb=20 70 540 810]{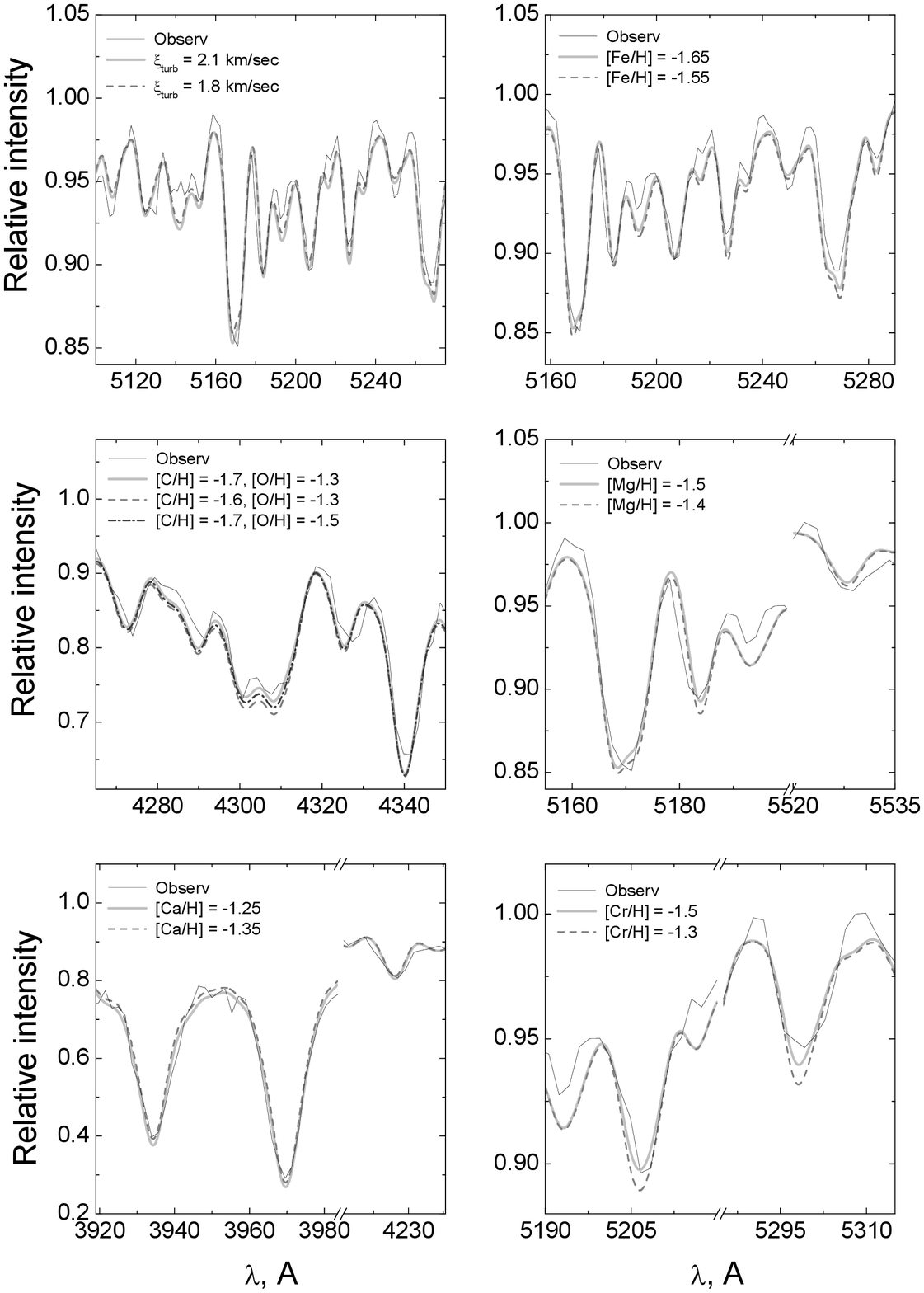}
\captionstyle{normal}
\caption{Determination of microturbulence velocity $\xi_{\rm
turb}$ and Fe, C, Mg, Ca, and Cr abundances for NGC\,6229. The
observed spectrum (solid lines) and theoretical spectra (dashed,
dashed-and-dotted, and dotted lines) calculated for different
[X/H] values with constant abundances of other elements (see
Sections 3.2, 4) are shown.} \label{fig7:Khamidullina_n}
\end{figure*}

\begin{figure*}[tbp!!!]
\setcaptionmargin{5mm} \onelinecaptionstrue
\includegraphics[angle=0,width=0.9\textwidth,bb=20 70 540 810]{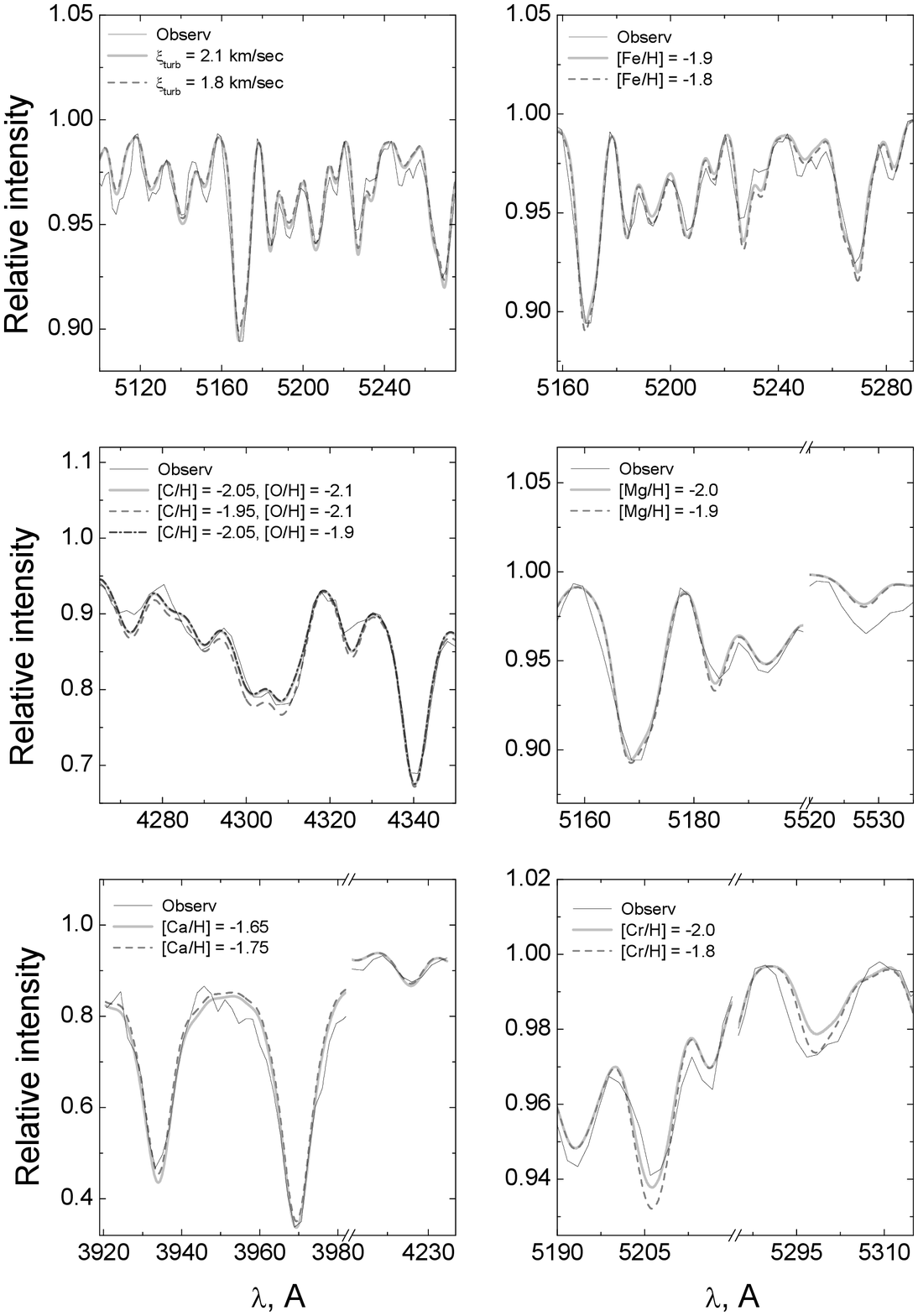}
\captionstyle{normal}
\caption{The same as in Fig.~\ref{fig7:Khamidullina_n}, but for
NGC\,6779.} \label{fig10:Khamidullina_n}
\end{figure*}

\begin{figure*}[tbp!!!]
\setcaptionmargin{5mm} \onelinecaptionstrue
\includegraphics[angle=0,width=0.9\textwidth,bb=20 70 540 810]{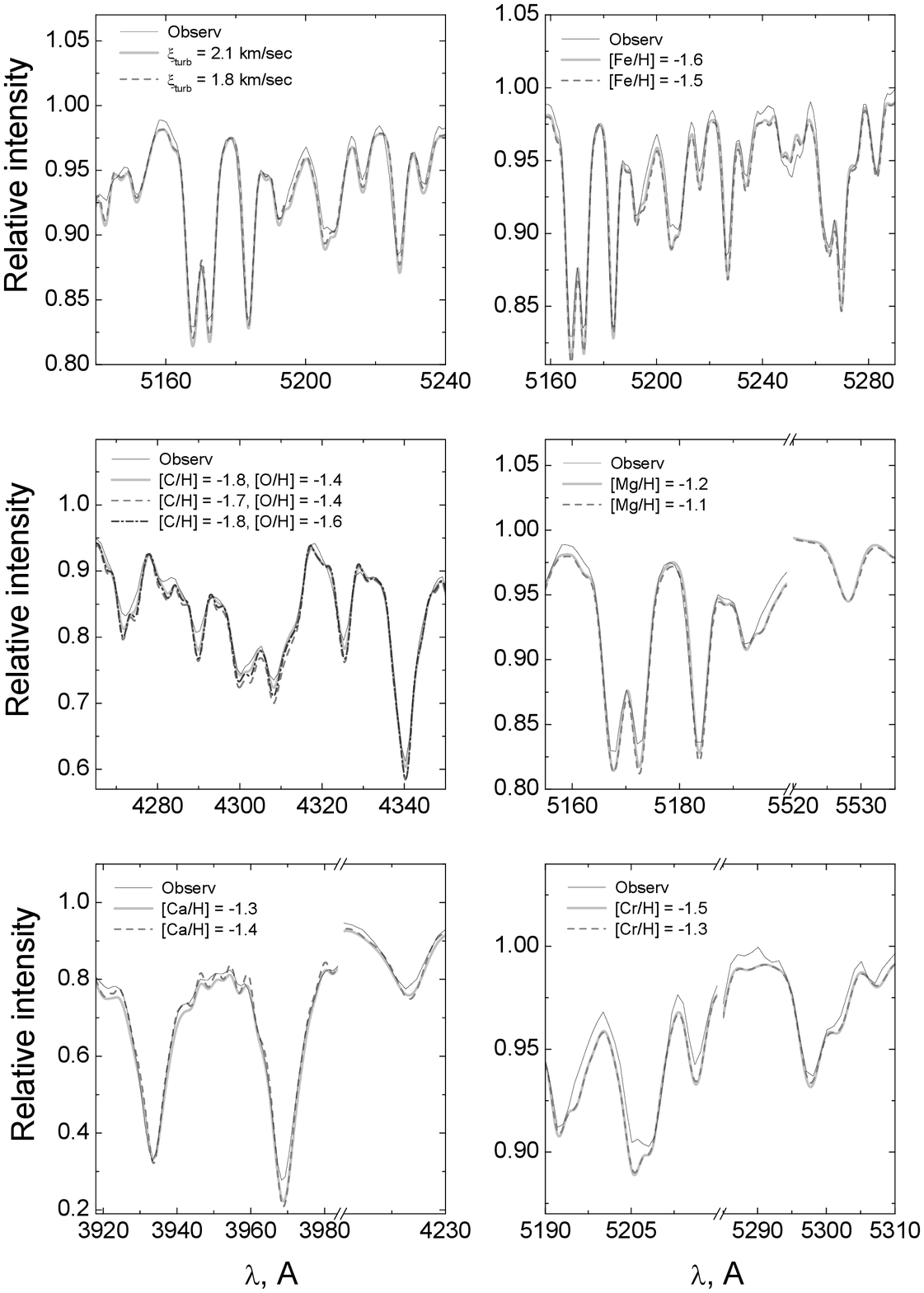}
\captionstyle{normal}
\caption{The same as in Fig.~\ref{fig7:Khamidullina_n}, but for
NGC\,5904.} \label{fig12:Khamidullina_n}
\end{figure*}

\begin{figure*}[tbp!!!]
\setcaptionmargin{5mm} \onelinecaptionstrue
\includegraphics[angle=0,width=0.9\textwidth,bb=20 70 540 810]{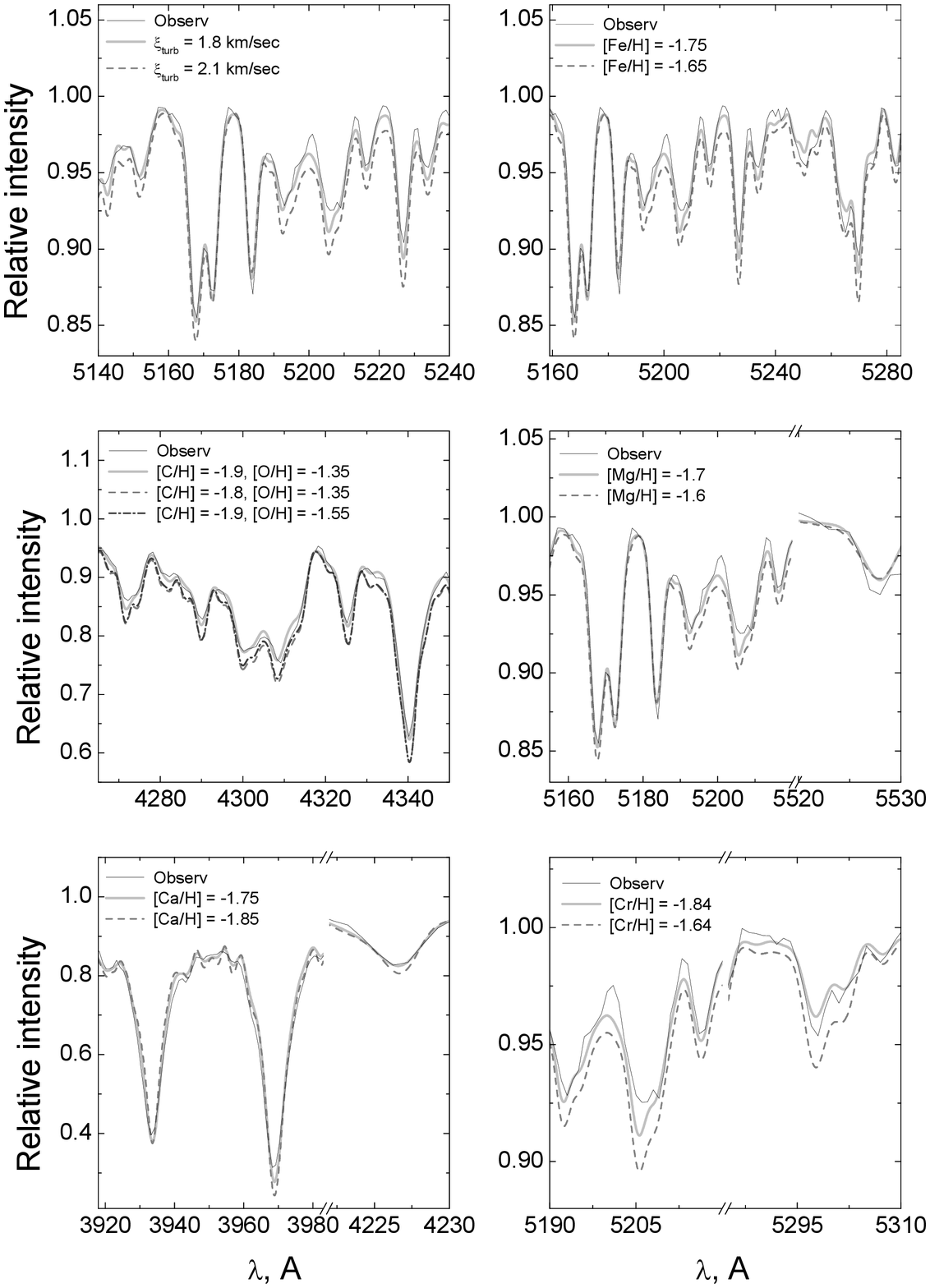}
\captionstyle{normal}
\caption{The same as in Fig.~\ref{fig7:Khamidullina_n}, but for
NGC\,6254.} \label{fig14:Khamidullina_n}
\end{figure*}

The number of stars with a given mass is calculated according to
the Chabrier mass function~\cite{chabrier:Khamidullina_n}. For
obtaining the synthetic spectra of plane-parallel, hydrostatic
stellar atmospheres for the derived iso\-chro\-nic parameters
($T_{\rm eff}$, $\log g$, [Fe/H]), we use the
CLUSTER~\cite{sharina:Khamidullina_n} software. Model atmospheres
are computed by interpolating the grids of Castelli and
Kurucz~\cite{cast1:Khamidullina_n} according to the technique
described in~\cite{sule2:Khamidullina_n}. For normalization we
model two spectra simultaneously: (1) with the allowance for
atomic lines and molecular bands in the spectral range covered, and
(2)~without it. Thus, the synthetic spectra are derived by
dividing~(1) by~(2). Note that during the modelling of all
spectra, a constant wavelength step of $\Delta\lambda =
0.05$~\AA was assumed, which ensures errors of the flux in
the continuum and in the lines no greater than $0.005\%$ of the
residual intensity. During the analysis of moderate-resolution
spectra (${\rm FWHM} \le 5$~\AA), it is possible to
investigate not only the individual lines but also broad blends
$\Delta \lambda \ge 5$~\AA\ consisting of lines and bands of many
atoms and molecules. Therefore, to determine the parameters and
abundances of GCs, it is necessary to achieve the best fit between
theoretical and observed spectra in the whole studied range. As
the analysis in~\cite{sharina:Khamidullina_n} with $S/N>100$
showed, this approach allows us to determine the cluster
parameters and the abundances of about 10 chemical elements. Our
abundances, determined by the strong Ca, Mg, Fe, CH lines and
blends, the majority of the Na, Al, Ba, Sr lines, and all the
lines with $\lambda
> 5300$~\AA, are differential, as the empirical oscillator strengths $gf$, derived in~\cite{shim11:Khamidullina_n} by the
solar spectrum analysis, are used. For the rest of the lines,
theoretical values of $gf$ are used which can cause, according to
the analysis of Shimanskaya et al.~\cite{shim11:Khamidullina_n}, a
systematic abundance underestimation of up to 0.07 dex. A common
microturbulence velocity $\xi_{\rm turb}$ for all the GC stars
(Table~3) is derived from the best fit between the strong and weak
iron lines in the theoretical and observed spectra.
Figures~\ref{fig7:Khamidullina_n}--\ref{fig14:Khamidullina_n}
show examples of Fe line variations at different $\xi_{\rm turb}$
values. Table~3 shows the $\xi_{\rm turb}$ values derived by us.
Note that the use of a common value of $\xi_{\rm turb}$ for all
the cluster stars is not quite proper. However, the analysis of
the integrated spectra of GCs does not allow us to derive a
dependence of $\xi_{\rm turb}$ on stellar parameters. Moreover,
the main contribution to the optical spectra is made by the
stars located at the MS turnoff point at~$T_{\rm eff}
=6200$--$7500$ K and by the brightest asymptotic branch giants.
For such stars, the microturbulence velocity is
within the range of $\xi_{\rm turb} = 1.8$--$2.4$ km\,s$^{-1}$.
That is why the use of a common value of $\xi_{\rm turb}$ cannot
cause any significant errors in the abundance of chemical
elements. The macroturbulence and rotation of individual cluster
stars are neglected in our calculations, as those effects amount
to less than $30$ km\,s$^{-1}$ and, consequently, are
insignificant at our resolution. For the same reason, we do not
consider the velocity dispersion of the stars within the cluster.

Fitting the continuum of the observed spectra to the theoretical
ones is conducted by the successive use of the following digital
filters (see the description of MIDAS{\footnote {\tt
http://www.eso.org/sci/software/esomidas/doc/\linebreak
/user/98NOV/volb/}} for details): (1)~determining the maximum spectral intensity within a specified wavelength range
($5\,{\rm FWHM}$), (2)~smoothing filter (running median) with a
radius of $6\,{\rm FWHM}$, where $\rm FWHM$ denotes the spectral
resolution of the spectrograph in \AA.

\begin{figure*}[tbp!!!]
\setcaptionmargin{5mm} \onelinecaptionstrue
\includegraphics[angle=-90,scale=0.6]{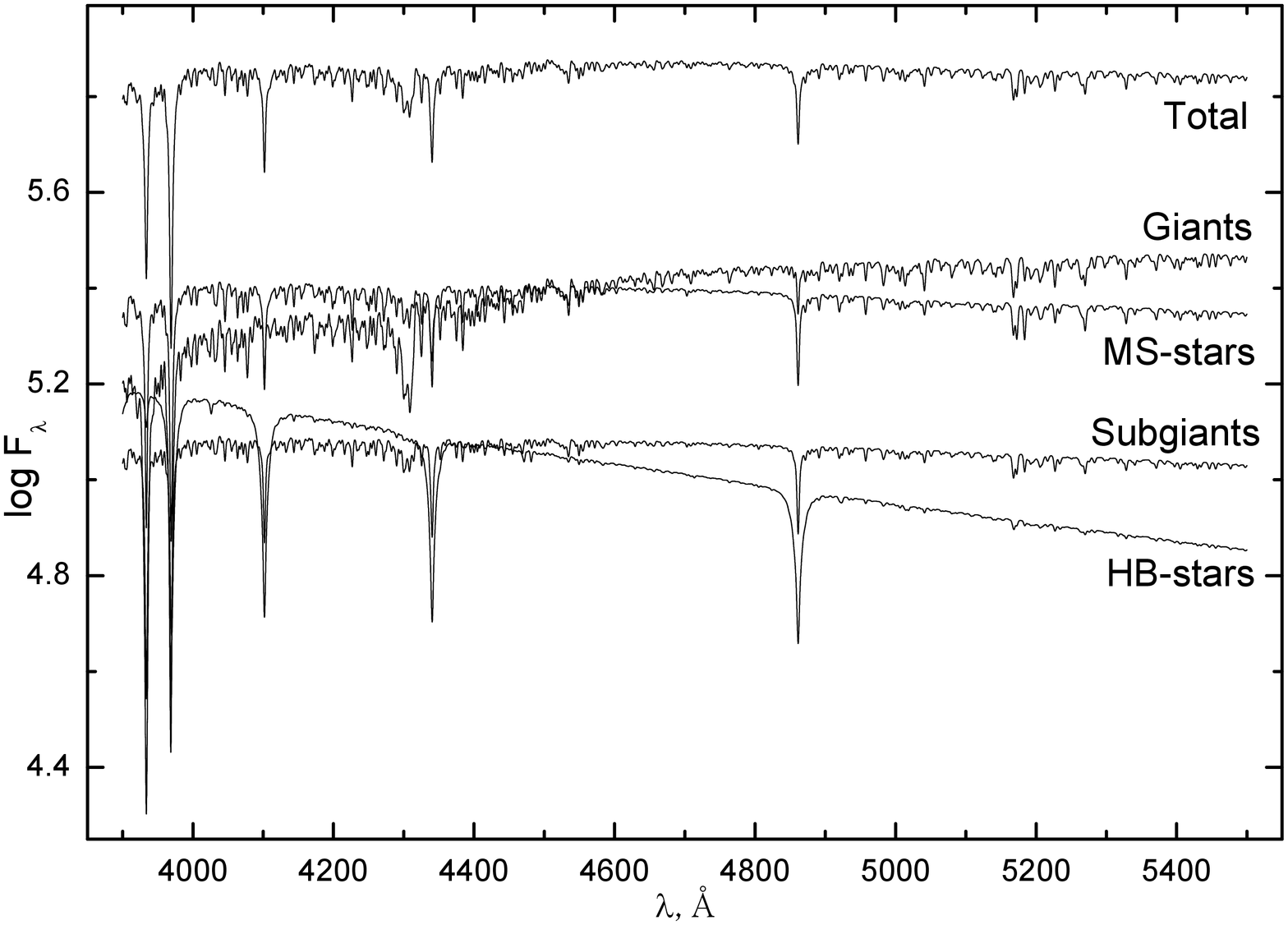}
 \vspace{-3mm}
\captionstyle{normal}
\caption{The contribution of different types of stars to the
integrated GC radiation. The example of NGC\,6229 is considered. }
\label{figstages:Khamidullina_n}
\end{figure*}

The contribution of the stars of different types to the integrated
GC flux is described by the example of NGC\,6229 in
Fig.~\ref{figstages:Khamidullina_n}. This cluster has an HB of an
intermediate type, although its blue and red ends are densely
populated with stars. Hot blue HB stars determine the width and
intensity of the hydrogen lines. These objects contribute up to
40$\%$ of the energy distribution in the blue region of the
investigated spectral range. Red giants, dominating in its red
region, are the main source of molecular bands and lines of
neutral atoms of heavy elements. Note that the molecular lines in
the sum spectrum of the stars in the red giant branch are much
more noticeable than in the spectrum of the MS and subgiants
branch stars. As a result, the integrated spectrum of the whole
cluster noticeably differs from the spectra of all its components.

\subsection{Abundance Determination}

Variations of [Fe/H], $\log~g$, and $T_{\rm eff}$ allow us to
calculate the spectra of various cluster stars. Summing up these
spectra taking into account the stellar mass functions produces the total synthetic spectrum of a GC.

The average chemical abundances in the atmospheres of GC stars are
determined by matching the observed and theoretical line
profiles, line blends, and molecular bands. As explained in
the previous paragraph, the resolution of our spectra does not
allow us to determine chemical abundances from individual
lines. In view of that, we adopted such abundances for each element for which
all the blends (including the lines of this element in the
studied wavelength interval) are best described by the theoretical
spectrum.

First, we determine the chemical abundance of the elements whose
lines and/or molecular bands dominate in the spectra which allows
us to obtain [X/H] values with an accuracy of $\Delta
[X/H] \sim 0.1$--$0.2$~dex. These elements are Fe, Ca, Mg,
C. The line profiles of Ti, Cr, Co, Mn,~N are analyzed by varying
their abundances using previously derived fixed abundances of the
elements from the first group. The accuracy of chemical abundance
estimation for the elements of the second group, the lines of
which are much weaker and more blended than the lines of the
elements from the first group, is about $0.2$--$0.3$~dex.
When we use moderate-resolution spectra, the individual elements
do not have any noticeable details in the observed wavelength
range, but they influence the molecular and ionization equilibrium
of other elements. Oxygen belongs to such elements; as its abundance increases, a part of atomic carbon participates in
the formation of the CO molecule which reduces the intensity of
the CN and CH molecular bands. The elements Al, Si, V, Ni
influence the electronic equilibrium in stellar atmospheres only
slightly, and also do not have observable lines at moderate
resolution. Na should also be included in this group, as its
resonance lines are heavily distorted by interstellar lines and
are unsuitable for analysis.

\section{RESULTS}

This is the first time that the chemical abundances for the clusters NGC 6229 and NGC 6779 have been derived. Tables~3--5 and
Figs.~\ref{fig7:Khamidullina_n}--\ref{fig14:Khamidullina_n} show
the final results for them and for the comparison clusters
NGC\,5904 and NGC\,6254. Table~3 summarizes the GC fundamental
parameters: age, helium abundance, iron [Fe/H] and
$\alpha$-element abundance averaged over the GC. Tables~4 and 5
show the abundances of individual chemical elements.
Figs.~\ref{fig5:Khamidullina_n}--\ref{fig6:Khamidullina_n} show
the results of the comparison of the observed and theoretical
H$\beta$, H$\gamma$, H$\delta$ line profiles, calculated based on
the isochrones with the parameters derived from the photometric
data (see above).
Figs.~\ref{fig7:Khamidullina_n}--\ref{fig14:Khamidullina_n} show
the profiles of the lines and molecular bands of different
chemical elements that were used to estimate their abundances. For
each element, two--three variants of the line profiles with
different abundances are presented for reliability assessment
and accuracy check. It is necessary to note that when fitting the
theoretical spectrum to the observed one, we assume that the
light from GC stars in the main evolutionary stages is
adequately represented in these two spectra. It is also assumed
that the observed spectrum is not distorted by background objects.
In reality, the last condition is not always met, especially in
the case of GCs close to the Galactic plane. Observations with
different slit orientations allow us to eliminate the distortions
caused by background stars in most cases. The spectra of test GCs
from Schiavon et al.~\cite{schiavon:Khamidullina_n} were derived
using the long-slit scanning method, and, thus, they may contain
such distortions.

As was mentioned in the Introduction, age, [Fe/H], and $Y$ were
estimated using deep photometry results from the
literature. Table~3 shows the summary of those estimates.
Generally, our results of the above-mentioned estimation of these
parameters, derived by means of cluster spectrum analysis, agree
with the ones from the literature. However, there are a number of
discrepancies. Our age estimate for the well studied GC NGC\,5904
exceeds the value derived in~\cite{vandenberg:Khamidullina_n} by
0.9~Gyr, and the helium abundance is excessive. These parameters
are determined by the H\,I line profile analysis, as explained in
Section 3.2. Our [Fe/H] values, derived using the spectroscopic
method, are in most cases smaller by 0.2--0.3 dex than those from
the literature, derived by the $C$--$M$ diagram analysis. It is
unlikely that these discrepancies are caused by background
stars caught in the slit. The projected vicinities of the
investigated clusters mainly contain stars of the Galactic disk
with close-to-solar metallicities. We have not detected any
emissions in the observed cluster spectra. Thus, the noted
differences are likely associated with the distinctions of the
applied methods and models. When analyzing the photometric data,
the errors of color indices and distances to the clusters, and
also the use of theoretical isochrones from different authors
produce random and systematic shifts in the derived
parameters~\cite{gallart:Khamidullina_n}. The use of different
atomic and molecular data sets, different model atmospheres for
the stars, and different principles of modelling simple stellar
populations leads to differences in the abundance estimates and cluster evolutionary parameters.

\begin{figure*}
\setcaptionmargin{5mm}
\onelinecaptionsfalse
\includegraphics[angle=-90,width=0.8\textwidth]{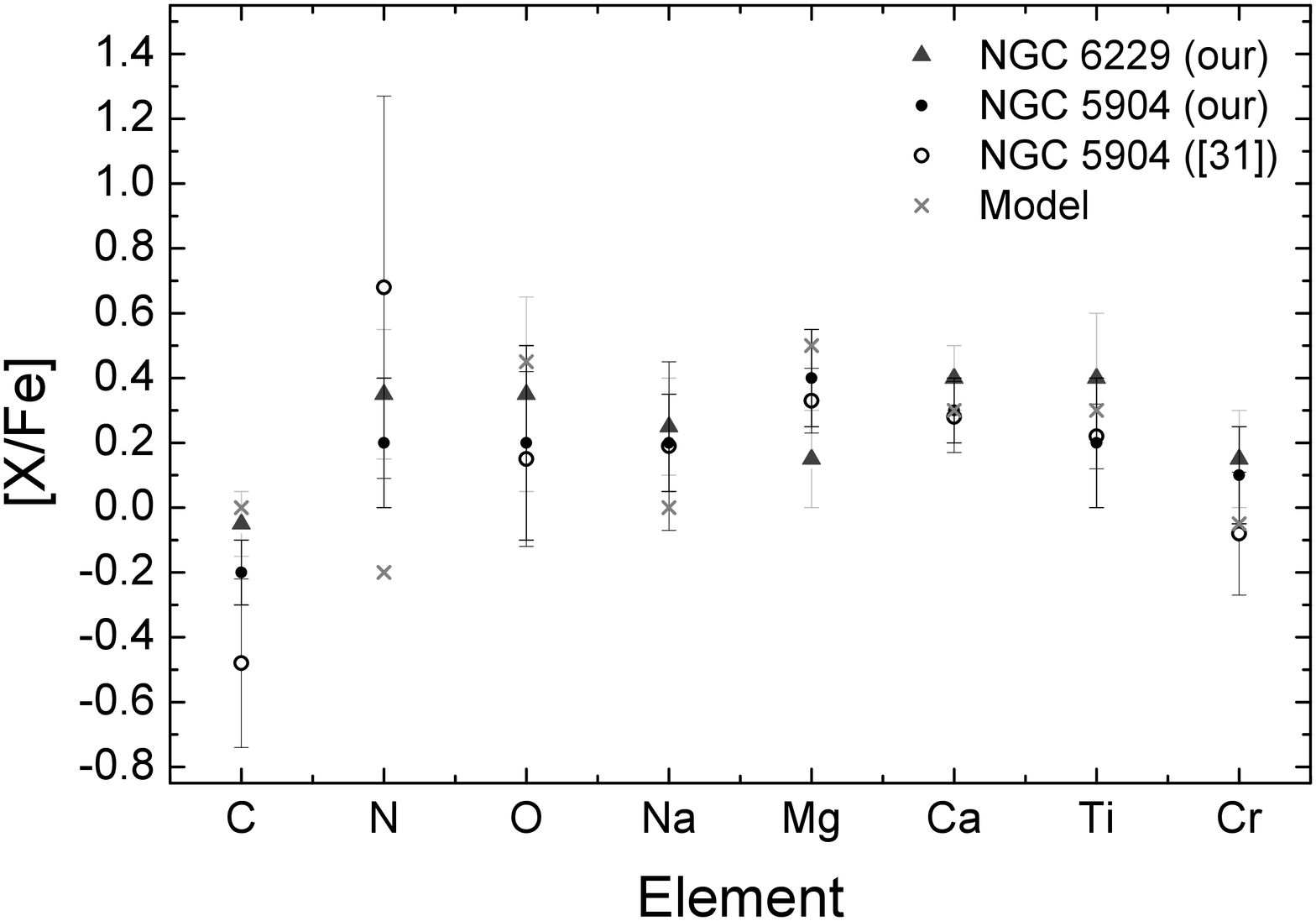}
\captionstyle{normal}
\caption{A comparison of the chemical abundances in NGC\,6229 and
NGC\,5904 according to the results of this paper. The open
circles denote the data from~\cite{roediger:Khamidullina_n}. }
\label{fig16:Khamidullina_n}
\end{figure*}

\begin{figure*}
\setcaptionmargin{5mm}
\onelinecaptionsfalse
\includegraphics[angle=-90,width=0.8\textwidth]{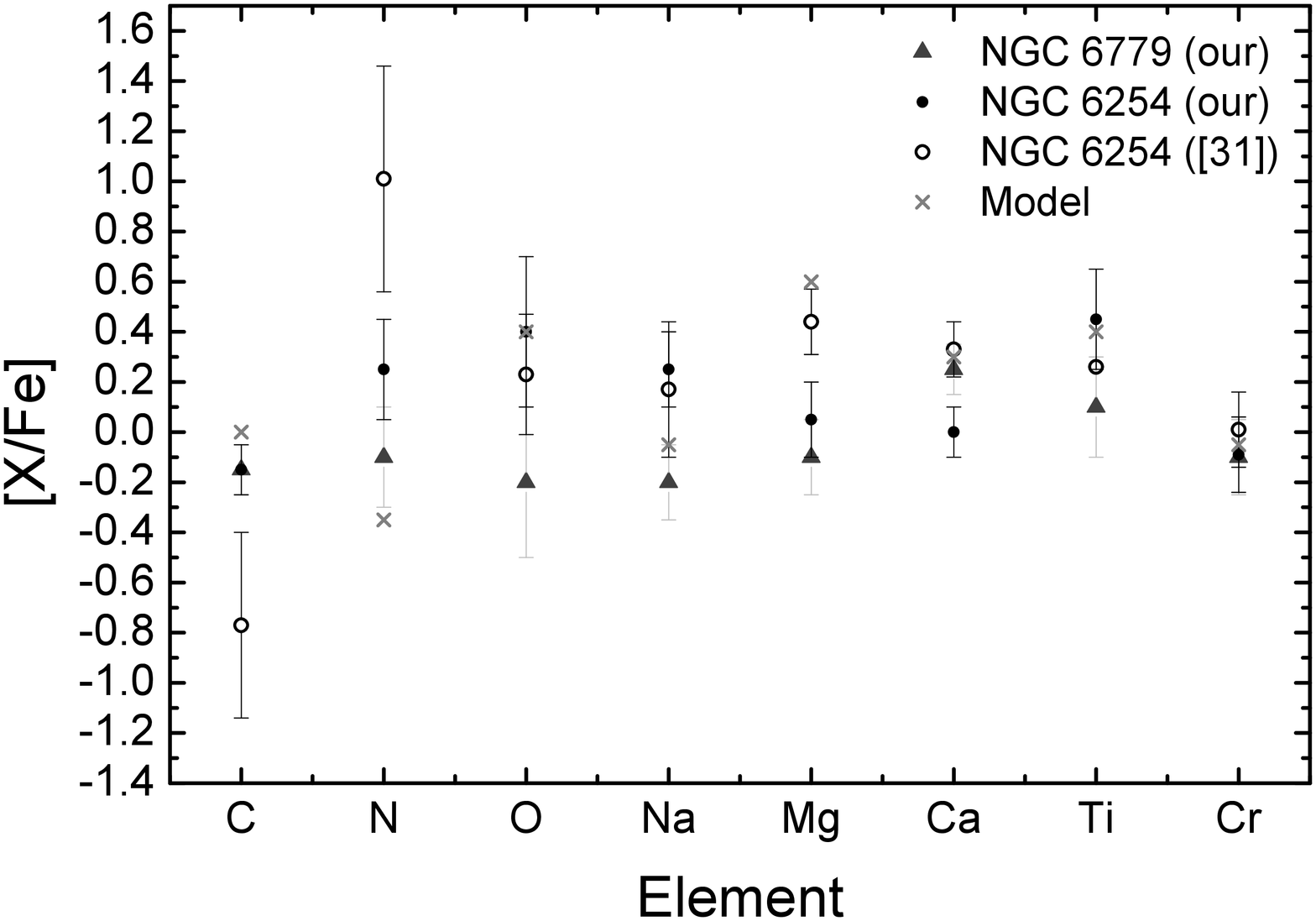}
\captionstyle{normal}
\caption{A comparison of the chemical abundances in NGC\,6779 and
NGC\,6254 according to the results of this paper. The open
circles denote the data from~\cite{roediger:Khamidullina_n}. }
\label{fig17:Khamidullina_n}
\end{figure*}

Tables 4 and 5, and
Figs.~\ref{fig16:Khamidullina_n}~and~\ref{fig17:Khamidullina_n}
show the comparison of our abundance estimation results for
different elements and the results
from~\cite{pritzl05:Khamidullina_n}
and~\cite{roediger:Khamidullina_n} for the well studied clusters
NGC\,5904 and NGC\,6254. The abundance differences are greater
than the given errors only for C~and~N. A possible explanation is
that we analyze the average chemical abundance of the whole
cluster and not only of the high-luminosity stars, as the majority
of the authors do. The products of the CNO cycle could be brought
up to the surface of these stars, and this may cause an increase
in the nitrogen abundance and a decrease of that of carbon. Note
that the considerable contribution of red giants to the cluster
spectrum probably increases the distortions of the estimated
C~and~N abundances, but their amplitude should be much smaller
than in the investigation of individual stars. There are also
significant differences in the Mg~and~Ca abundances for NGC\,6254.
Considerable variations of C, N, O, Mg, Na, and Al from star to
star were found for a number of massive Galactic
GCs~\cite{gratton04:Khamidullina_n}. In one and the same cluster,
one can often observe two or more stellar populations with
different chemical abundances forming different sequences in the
$C$--$M$ diagram. This is explained by the mixing of matter
after the CNO cycle. The Mg abundance variations after such
processes are expected to be around $\Delta Mg \le
0.3$ dex~\cite{pritzl05:Khamidullina_n}. Note once again that the
estimates from the literature are based mainly on the
investigations of red giants, which are not the true indicators of
GC initial abundances because of the mixing of matter, to which
their atmospheres are liable (the first dredge-up). Therefore,
more detailed theoretical and observational investigations are
necessary in order to check how much the cluster abundance
variations depend on the physical characteristics of the studied
stars.

Almost all our estimated abundances agree with the theoretical
calculations of the models of the chemical evolution of the Galaxy and also
with the average values for dwarfs at the given metallicity
(see~\cite{samland:Khamidullina_n,alibes:Khamidullina_n,kobayashi:Khamidullina_n}).
However, the scatter of the observed values, with which model
characteristics are compared, is often large.

Carbon is produced in the triple $\alpha$-process with hydrostatic He
burning in stars of various masses. The average abundance of C
in halo and disk dwarfs varies little with time (does not depend on
[Fe/H]): $[C/Fe] \sim 0$~\cite{alibes:Khamidullina_n}. We
have derived similar values.

The nitrogen abundance decreases with metallicity as
massive stars do not produce primary N. With the [Fe/H] increase,
the production of the so-called secondary N by moderate-mass stars
increases until the Fe production by Ia supernovae compensates
this process at $[Fe/H] \sim -1$. The theoretical [N/Fe]
values vary from $0$~to~$-0.5$ in the following range:
$ -2 < [Fe/H] < 1$. Our estimates are closer to the
theoretical ones than the data from the literature for the giants
in NGC\,5904 and NGC\,6254.

Oxygen, as the majority of other $\alpha$-elements, is produced by
massive stars only. $[O/Fe] \sim 0.5$ at $[Fe/H] \le
1$. Our values correspond quite well to those calculations, except
for NGC\,6779, for which we obtained lower abundance estimates for
the other $\alpha$-ele\-ments as well. There were no
emissions or noticeable distortions from background stars detected
in the spectra. However, one cannot exclude the latter possibility
for NGC\,6779, which is closer to the plane of the Milky Way than
the other three clusters.

The evolution of the abundance of magnesium as a function of metallicity is
impossible to explain only by its formation in massive stars
(\cite{alibes:Khamidullina_n} and references therein).
The theoretical [Mg/Fe] values in the range of $ -2 < [Fe/H] <
1$ are usually a little lower than the observed ones. Our results
agree better with the theoretical values. However, the scatter of the observed data is large. On the
whole, one can conclude that the [Mg/Fe] values for the selected
clusters except NGC\,6779 are within the normal range.

The value of [Ca/Fe] is in close agreement with theory for all the
selected clusters except NGC\,6254. We estimated the abundance of
this element not only from the Ca\,I $\lambda~4226$~\AA\ resonance
line but also from the H and K Ca\,II lines, the profiles of which
vary greatly depending on the selected isochrone, as these
profiles are determined not only by the Ca content but also by
age, $Y$, and [Fe/H]. Thus, an error in the [Ca/Fe] estimate is
possible, but much less probable than the possible errors for the
elements heavier than Ca, which have no broad spectral features.

Tables~4~and~5 and
Figs.~\ref{fig16:Khamidullina_n}~and~\ref{fig17:Khamidullina_n}
show that, on the whole, the chemical abundances for NGC\,5904 and
NGC\,6254 are similar to the ones for NGC\,6229 and NGC\,6779. The
published values of [Fe/H] for NGC\,6779 and NGC\,6254 are smaller
than the same values for NGC\,6229 and NGC\,5904. This fact, in
view of the similarity of the horizontal branches of NGC\,6229 and
NGC\,5904, and also of NGC\,6779 and NGC\,6254, has been a reason for associating the clusters of higher metallicity with the young Galactic
halo, and clusters of lower metallicity---with the old
one~\cite{mars:Khamidullina_n}. Following our results (see
Table~3), there are no two objects out of the four investigated
that are totally similar in all parameters. The similarity in the
chemical abundances is indicative of the fact that they belong to
the old Galactic halo, in which nucleosynthesis proceeded
according to the main evolutionary stages under the influence of
SN\,II and SN\,Ia supernovae. In order to associate the clusters
with one or another Galactic subsystem, a more detailed analysis of
the chemical, kinematic, and structural characteristics is
necessary.

\section{CONCLUSIONS AND SUMMARY}

The method we used for modelling and analysis of the integrated
spectra of the clusters was previously discussed
in~\cite{sharina:Khamidullina_n}. In this paper we expanded it by
including the investigation of the cluster $C$--$M$
diagrams and their detailed comparison with the theoretical
isochrones. Now, as a result, this method combines the
possibilities of using both the spectroscopic, integrated along
the slit data, and photometric data on the stellar populations of
the clusters. We use the results of deep stellar photometry from
the literature to select an isochrone that best fits the cluster
$C$--$M$ diagram. The summing-up of the stellar synthetic
blanketed spectra, calculated using model atmospheres, is
conducted using the stellar luminosity function of
Chabrier~\cite{chabrier:Khamidullina_n}. The matching of the
theoretical and observed cluster spectra is carried out
iteratively. Initially, the distribution of cluster stars by mass,
radius, and $\log~g$ is preset using the literature data on the
theoretical isochrone which fits best the observed $C$--$M$
diagram. For the specification of stellar ages and helium
abundances, the profiles of the hydrogen lines are analyzed,
because they practically do not depend on other parameters. The
abundance of iron is then varied, the lines of which prevail in
the optical spectrum even at a low metallicity. Upon achieving the
closest agreement in these three parameters, the abundances of the
$\alpha$-elements Mg,~Ca,~and~C are derived. The molecular
bands and lines of these elements are the most significant in the
spectrum. For the determination of the abundances of other
elements, complex blends consisting of many lines are used. As a
result, the abundances of about 10~elements are estimated. Note
that the use of a considerable number of lines with empirical $gf$
values allows us to obtain differential abundances (i.e., bound to
the $gf$ values of solar spectral lines) of Ca, Mg, Fe, C, and, in
some cases, other elements.

In this paper, we determined the ages, specific helium abundances
$Y$, and abundances of Fe, C, N, O, Na, Mg, Ca, Ti, and Cr for the
following four Galactic GCs using the developed method: NGC\,6229,
NGC\,6779, NGC\,5904, NGC\,6254. The chemical abundances for the
first two objects were estimated with a spectroscopic method for
the first time. According to our estimates, these GCs turned out
to be about 1~Gyr younger than NGC\,5904 and NGC\,6254, which are
accepted in the literature as their analogues.  The helium
abundance that we determined for the last two clusters is higher
($Y=0.30$) than the one in the literature. The clusters
NGC\,6229 and NGC\,6779 show the values $Y = 0.23$ and
$Y = 0.26$ respectively, which are closer to the value of
the primordial helium abundance in the Universe $Y =
0.25$~\cite{komatsu11:Khamidullina_n}. The
$\alpha$-element abundances\linebreak in the metal-poor
($[Fe/H] \sim -1.6$) NGC\,6229\linebreak and
NGC\,5904 turned out to be high\linebreak
($[\alpha/Fe] = 0.28$--$0.35$) as in the old massive
globular clusters of the Galaxy; massive type II supernovae
contributed greatly to their chemical enrichment. However, the
$\alpha$-element abundances in the investigated clusters with low
central density, NGC\,6779 and NGC\,6254, turned out to be quite
low ($[\alpha/Fe] = 0.08, 0.025$), while their
metallicity was low ([$Fe/H] = -1.9, -1.7$). Such
$[\alpha/Fe]$ values are not typical for low metallicity
Galactic clusters. For this reason, the search for a dependence of
the chemical abundance of GCs on their structure and kinematic
characteristics can be the subject of future investigations. A
comparison of the derived abundances and the literature values for
the well studied GCs NGC\,5904 and NGC\,6254 showed that our
measurements are consistent with the corresponding average
estimates from the literature for cluster giants with the average
accuracy of about $0.15$~dex, except for C and N. This is probably due
to the difference between the average chemical
composition in the stellar atmospheres for the whole cluster,
derived in this paper, and the chemical composition of the
atmospheres of high-luminosity stars listed in the literature. The products of the
CNO cycle can be transported to the surface of such stars, and
their study does not provide accurate information on the primary
chemical abundance of GCs.

The values of [Fe/H], $Y$, and age which we derived by
spectroscopic analysis allow us to evaluate the true correlation
between the observed $C$--$M$ diagram and the corresponding theoretical
isochrone. At present, there is no possibility of combining the deep stellar photometry data with the
spectroscopic methods of estimating the element abundances and
evolutionary parameters when analyzing the
characteristics of stellar populations in extragalactic GCs. Thus, our investigation provides
essential data for understanding the characteristics of GCs in
remote galaxies.

\begin{acknowledgments}
We acknowledge attribution of the RFBR regional grant number
14-02-96501-r-ug-a. The financial support, however, was not
fully received because of the Karachay-Cherkess Republic.
V.~V.~Shimansky thanks the RFBR (grant~13-02-00351) for the
financial support, and is also grateful for funding from the
subsidy sponsored under the state support of the Kazan (Volga
region) Federal University for the purpose of competitive recovery
among the world's leading Research and Education Centers. This
research has made use of SAO/NASA~ADS, and the SIMBAD database,
operated at CDS, Strasbourg, France. We are grateful to the
anonymous reviewer for the helpful advice, which allowed us to
improve the paper.
\end{acknowledgments}

\onecolumngrid

\begin{flushright}
{\it Translated by N.~Oborina}
\end{flushright}

\enddocument